\documentclass[reprint,amsmath,amssymb,aps,prc,longbibliography,lengthcheck]{revtex4-1}
\usepackage{graphicx}
\usepackage{amsmath}
\usepackage{amssymb}
\usepackage{ulem}

\usepackage{epsfig}
 \usepackage{lineno}

\begin{document}

\title{Effects of entrance channels on the deexcitation properties
of the same compound nucleus formed by different pairs of
collision partners}

\author{G. Mandaglio$^{1,2}$}\thanks{gmandaglio@unime.it}
\author{ A. Anastasi$^{3}$, F. Curciarello$^{4}$, G. Fazio$^{3}$, G. Giardina$^{3}$}
\affiliation{$^{1}$ Dipartimento di Scienze Chimiche, Biologiche, Farmaceutiche ed Ambientali, University of Messina, Messina, Italy}
\affiliation{$^{2}$ INFN Sezione di Catania, Catania, Italy}
\affiliation{$^{3}$ Dipartimento di Scienze Matematiche e Informatiche, Scienze Fisiche e Scienze della Terra, University of Messina, Messina, Italy}
\affiliation{$^{4}$ INFN Laboratori Nazionali di Frascati, Frascati, Italy}
\author{A. K. Nasirov}\thanks{nasirov@jinr.ru}
\affiliation{
JINR - Bogoliubov Laboratory of Theoretical Physics, Dubna, Russia}
\affiliation{
Institute of Nuclear Physics, Academy of Science of Uzbekistan, Tashkent, Uzbekistan}

\begin{abstract}
The   properties of deexcitation of the same $^{220}$Th compound nucleus (CN) formed by  different mass  (charge)  asymmetric
 reactions are investigated.  It is demonstrated that the effective  fission barrier $<B_{\rm fis}>$ value  being a function of the   excitation energy $E^*_{\rm CN}$  is strongly sensitive to the various orbital angular momentum $L=\ell\hbar$ distributions of CN formed with the same excitation energy 
 $E^*_{CN}$ by the  very different entrance channels $^{16}$O+$^{204}$Pb,  $^{40}$Ar+$^{180}$Hf, $^{82}$Se+$^{138}$Ba and $^{96}$Zr+$^{124}$Sn.
  Consequently, the competition between the fission and evaporation of light particles (neutron, proton, and $\alpha$-particle) processes along the deexcitation cascade of CN  depends
  on  the orbital angular momentum distribution of CN. Therefore, the ratio between the evaporation residue cross sections obtained
   after emission of  neutral and charged particles  and  neutrons
   only for the same CN with a given excitation energy $E^*_{CN}$ is
   sensitive to the mass  (charge)  asymmetry of reactants in  the entrance channel.
\end{abstract}


\maketitle

\section{Introduction}
It is well known to the scientific community that in heavy ion collisions, at low energies, the complexity of processes preceding the formation of reaction products strongly influences  their properties and nature, 
and that due to very transient characteristics of  these processes it is impossible to observe how they occur. Moreover, in the reactions there are products that are strongly determined by the first stage of the collision between the projectile and target nuclei leading to the capture of reactants and then to the evolution of the dinuclear system (DNS) \cite{volk} up to the formation of  products of the quasifission process in competition with the ones of 
the complete fusion process. In this last case, the complete fusion stage can lead to the fast fission (FF) products \cite{NPA17} (caused for angular momentum values $\ell>\ell_{\rm cr}$ since in this intervall of $\ell$ the fission barrier $B_{\rm fis}$ is zero, and the deformed mononucleus breaks down into two fragments without to reach the compound nucleus (CN) stage), and to reaction products caused by the deexcitation of CN leading to the formation of fusion-fission fragments in competition with the evaporation residue (ER) nuclei reached after  light particle emissions surviving fission \cite{PRC91,NPA17,FazioJP77,aglio2009,aglio2012} at each step of the cascade. In this complex context,  many ER nuclei can not be detected and identified due to
 concrete limits of experimental apparata and/or analysis of data. Therefore, in the analysis of experimental data there are unavoidable uncertainties on the identification and separation of the products that are formed in each step of the reaction process. Of course, also in calculation of the theoretical models there are serious uncertainties on the obtained results due to the assumptions made in the procedures and use of phenomenological models. In this paper, we present a detailed analysis and comparison of calculated results obtained by the study of the   $^{16}$O+$^{204}$Pb,  $^{40}$Ar+$^{180}$Hf, $^{82}$Se+$^{138}$Ba, and $^{96}$Zr+$^{124}$Sn very different mass asymmetry reactions (with mass asymmetry parameter $\eta$=0.86, 0.64, 0.26, and 0.13, respectively) leading to the same $^{220}$Th CN.  These reactions are characterized by various  values of the threshold excitation energy  $E^*_{\rm CN}$  
 due to different reaction barrier energies\cite{PRC91} of about 27.8, 35.5, 12, and 17.5 MeV, respectively, caused by different values of repulsive Coulomb and centrifugal rotational potentials  and the attractive nuclear interaction. 
 Since the reaction products formed  at deexcitation cascade of CN are the evaporation residues
  after neutrons and charged particle emissions  in competition with the fission process, it is interesting
  to analyze the effects related to the various  deexcitation modes of the same  CN with the same excitation energy $E^*_{\rm CN}$, since the formed CN is characterized by a specific  angular momentum distribution due to
 the  different mass (charge) asymmetry  in the entrance channel.

\section{Method}

The study of heavy ion collisions near the Coulomb barrier energies is based on  calculations
of the incoming path of projectile nucleus and
 finding the capture probability, taking into account the possibility of interaction
with different orientation angles of the axial symmetry axis of deformed
nuclei \cite{FazioJP77}.
 Moreover, the surface vibration of
the  nuclei, which are spherical in the ground state and deformed shape in
 the first excited $2^+$ state, is taken into consideration. The final results
are averaged over all orientation angles the axial symmetry axis of deformed
nuclei or vibrational states of the spherical nuclei. These procedures
are presented in the Appendix A and Appendix B  of the papers \cite{PRC91,NPA17}.

The capture of the projectile by the target is characterized
by the full momentum transfer of the relative momentum into
the intrinsic degrees
of freedom and shape deformation. The capture occurs if the following
necessary and sufficient conditions are satisfied.
The necessary condition of capture is
 overcoming the Coulomb barrier by projectile nucleus to be trapped
in the potential well of the potential energy surface (PES).
The collision dynamics is calculated by the solution of the equations of the relative distance $R$ and angular momentum $\ell$\cite{PRC91,NPA17}.

The condition of sufficiency for capture is the
 decrease of the relative  kinetic energy due to dissipation by friction
 forces up to values lower than  the depth of the potential well
 \cite{PRC72,NuclPhys05,FazioMPL2005}. The potential well is formed due to the competition of the short range nuclear attractive and
 the Coulomb and centrifugal repulsive potentials.
 This condition depends on the values of the beam energy and orbital angular momentum,
 the size of the potential well and intensity of
 the friction forces that cause dissipation of the kinetic energy
of the relative motions to internal energy of two nuclei.
So, the trapping of the collision path in the well means that the capture has occurred and the DNS is formed.
The lifetime of the DNS is determined by its
excitation energy $E^*_{\rm DNS}$ and by the size of the potential well. The
height of the inner barrier of the potential well is called the
quasifission barrier $B_{\rm qf}$ in our approach. This definition is related
to the quasifission process: in this case the DNS decays without reaching
the equilibrated shape of a compound nucleus \cite{NuclPhys05,FazioMPL2005,back}. The
alternative to the quasifission process in the evolution of  DNS is
the complete fusion of its constituent fragments. According to this
scenario the partial capture cross section $\sigma^{\ell}_{\rm cap}$ for a given relative energy
in the center-of-mass system $E_{\rm c.m.}$ and angular momentum value $\ell$
is the sum of the partial complete fusion $\sigma^{\ell}_{\rm fus}$ and quasifission $\sigma^{\ell}_{\rm qf}$ cross sections  \cite{PLB2010};

\begin{eqnarray}
\sigma_{\rm cap}^{\ell}(E_{\rm c.m.},\ell;\alpha_1,\alpha_2)&=& \sigma_{\rm fus}^{\ell}(E_{\rm c.m.},\ell;\alpha_1,\alpha_2)\nonumber\\ &+&\sigma_{\rm qf}^{\ell}(E_{\rm c.m.},\ell;\alpha_1,\alpha_2).
\label{eqcap}
\end{eqnarray}

The capture cross section is determined by the number
of partial waves which lead to the path of the total energy of
colliding nuclei to be trapped in the well of the nucleus-nucleus
potential after dissipation of a sufficient part of the initial
kinetic energy. The size of the potential well decreases with
increasing orbital angular momentum $\ell$.
Therefore,
the capture cross section is calculated by the formula
\begin{eqnarray}
\sigma_{\rm cap}(E_{\rm c.m.};\alpha_1,\alpha_2) & = &\sum^{\ell_d(E_{\rm c.m.})}_{\ell=0} \sigma_{\rm cap}^{\ell}(E_{\rm c.m.},\ell;\alpha_1,\alpha_2)\nonumber \\
& =& \frac{\lambda^2}{4\pi}\sum^{\ell_d(E_{\rm c.m.})}_{\ell=0}(2\ell+1)\nonumber\\
& \times&\mathcal{P}^{\ell}_{\rm cap}(E_{\rm c.m.},\ell;\alpha_1,\alpha_2),
\label{eqcapsum}
\end{eqnarray}
the complete fusion (CF) cross section is obtained as
\begin{eqnarray}
\sigma_{\rm CF}(E_{\rm c.m.};\alpha_1,\alpha_2)&=&\sum^{\ell_d(E_{\rm c.m.})}_{\ell=0} \sigma_{\rm cap}^{\ell}(E_{\rm c.m.},\ell;\alpha_1,\alpha_2)\nonumber\\ &\times& P_{CF}^{\ell}(E_{\rm c.m.},\ell;\alpha_1,\alpha_2),
\label{eqfus2}
\end{eqnarray}
and the quasifission cross section at the stage of DNS
 is obtained as
\begin{eqnarray}
\sigma_{\rm qf}(E_{\rm c.m.};\alpha_1,\alpha_2)&=&\sum^{\ell_d(E_{\rm c.m.})}_{\ell=0} \sigma_{\rm cap}^{\ell}(E_{\rm c.m.},\ell;\alpha_1,\alpha_2)\nonumber \\ &\times& [1-P_{CF}^{\ell}(E_{\rm c.m.},\ell;\alpha_1,\alpha_2)];
\label{eqqfis}
\end{eqnarray}
for details see \cite{NPA17}.
Therefore, the part of complete fusion reaching the compound nucleus formation is
\begin{eqnarray}
\sigma_{\rm fus}(E_{\rm c.m.};\alpha_1,\alpha_2)&=&\sum^{\ell_{\rm cr}}_{\ell=0} \sigma_{\rm cap}^{\ell}(E_{\rm c.m.},\ell;\alpha_1,\alpha_2)\nonumber\\ &\times& P_{CF}^{\ell}(E_{\rm c.m.},\ell;\alpha_1,\alpha_2),
\label{eqfus}
\end{eqnarray}
while the part going in fast fission is
\begin{eqnarray}
\sigma_{\rm ff}(E_{\rm c.m.};\alpha_1,\alpha_2)&=&\sum^{\ell_{\rm d}(E_{\rm c.m.})}_{\ell=\ell_{\rm cr}} \sigma_{\rm cap}^{\ell}(E_{\rm c.m.},\ell;\alpha_1,\alpha_2)\nonumber\\ &\times& P_{CF}^{\ell}(E_{\rm c.m.},\ell;\alpha_1,\alpha_2).
\label{eqff}
\end{eqnarray}

In relation (\ref{eqcapsum}) $\lambda$ is the de Broglie wavelength of the entrance channel
and $\mathcal{P}^{\ell}_{\rm cap}(E_{\rm c.m.},\ell;\alpha_1,\alpha_2)$ is the capture probability which
depends on the collision dynamics: $\mathcal{P}_{\rm cap}^{\ell}$ is 1 at $\ell_{\rm min}\leq\ell\leq\ell_d$, while is 0 if $\ell<\ell_{\rm min}$ or $\ell>\ell_d$.

That means the $\ell$ possible values leading to capture can form in some cases a ``window'' of angular momentum values because the friction coefficient is not so strong
 to trap the projectile in the potential well. Moreover, the maximal
value of partial waves ($\ell_d$) leading to capture is calculated
by the solution of the equations of the relative motion of
nuclei \cite{PRC72,NuclPhys05,FazioJP72}, and $\ell_{min}$ is the minimal value of  $\ell$ leading to capture.
In relation (\ref{eqfus2}), $P_{\rm CF}^\ell$ represents the complete fusion probability that the excited DNS -through the exchange of nucleons between the two interacting nuclei- evolves towards the complete fusion in competition the quasifission process that instead consists in the separation of the two constituent nuclei. The fusion probability is strongly dependent on the values of the intrinsic fusion barrier $B^*_{\rm fus}$ and quasifission barrier $B^*_{\rm qf}$, both sensitive to some parameters of the reaction.
For any details see  \cite{PRC91,NPA17,utamur}.

\section{Role of the angular momentum on the CN formation and its deexcitation}

\begin{figure}
\centering
\includegraphics[scale=0.35]{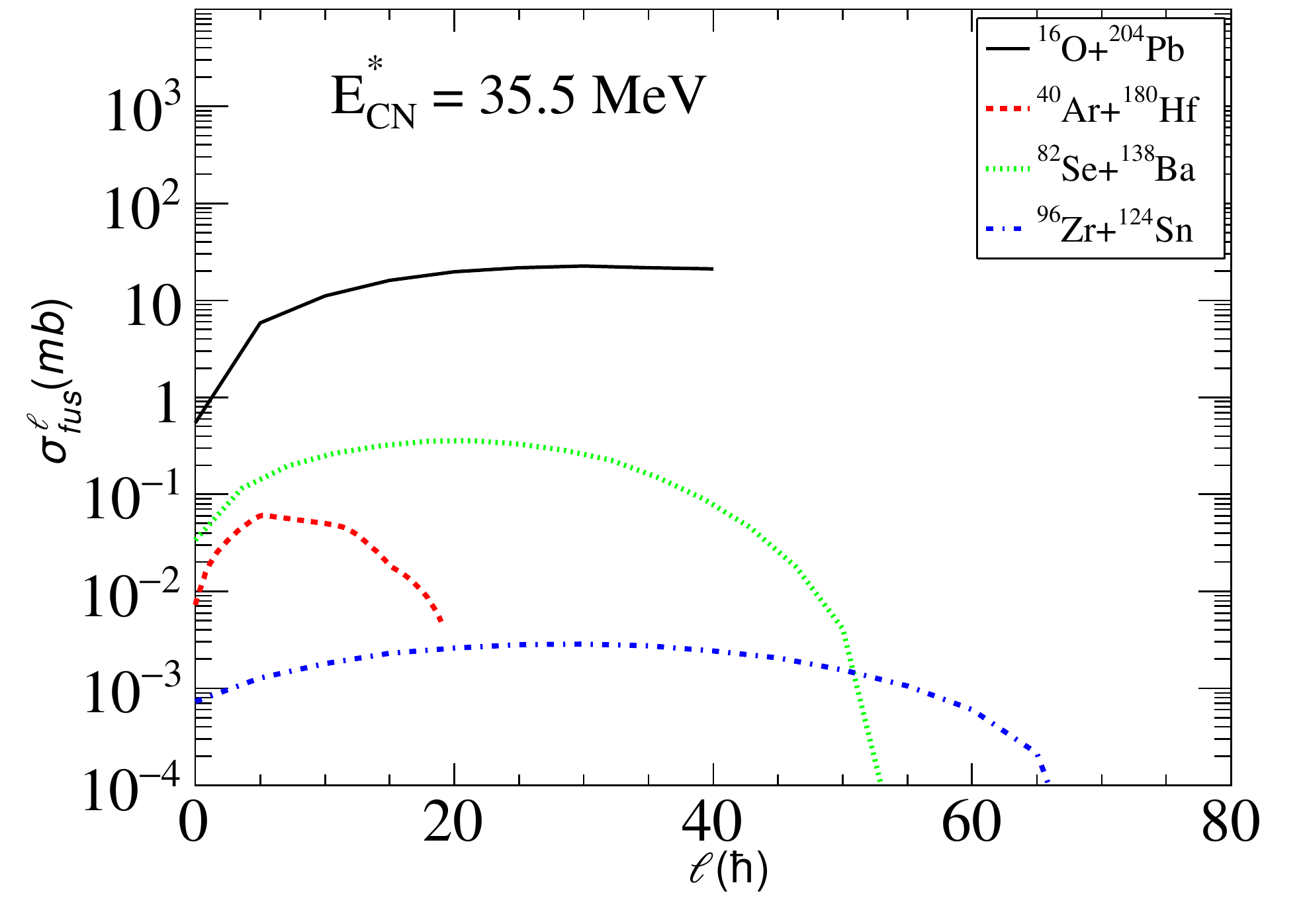}
\includegraphics[scale=0.35]{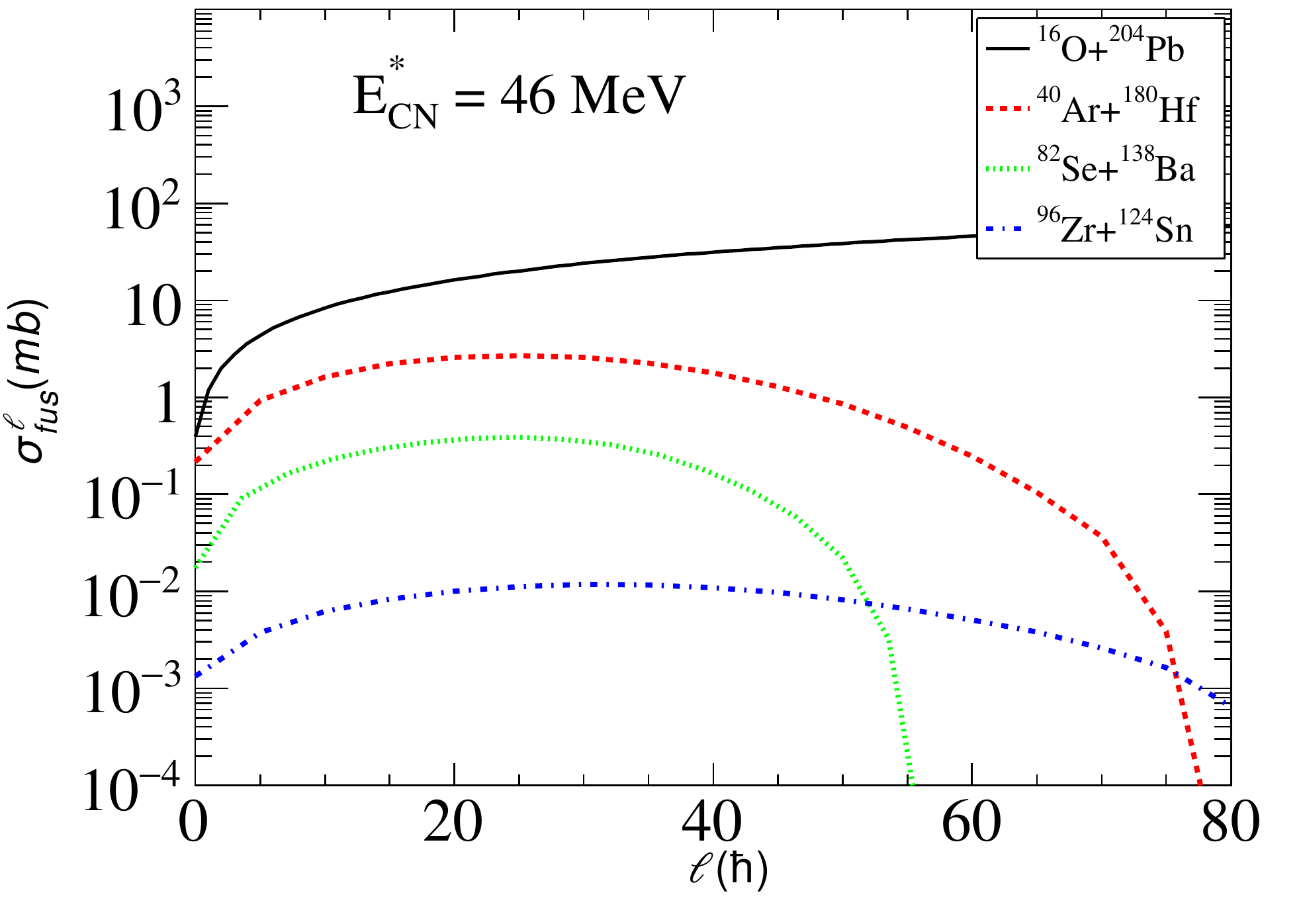}\\
\includegraphics[scale=0.35]{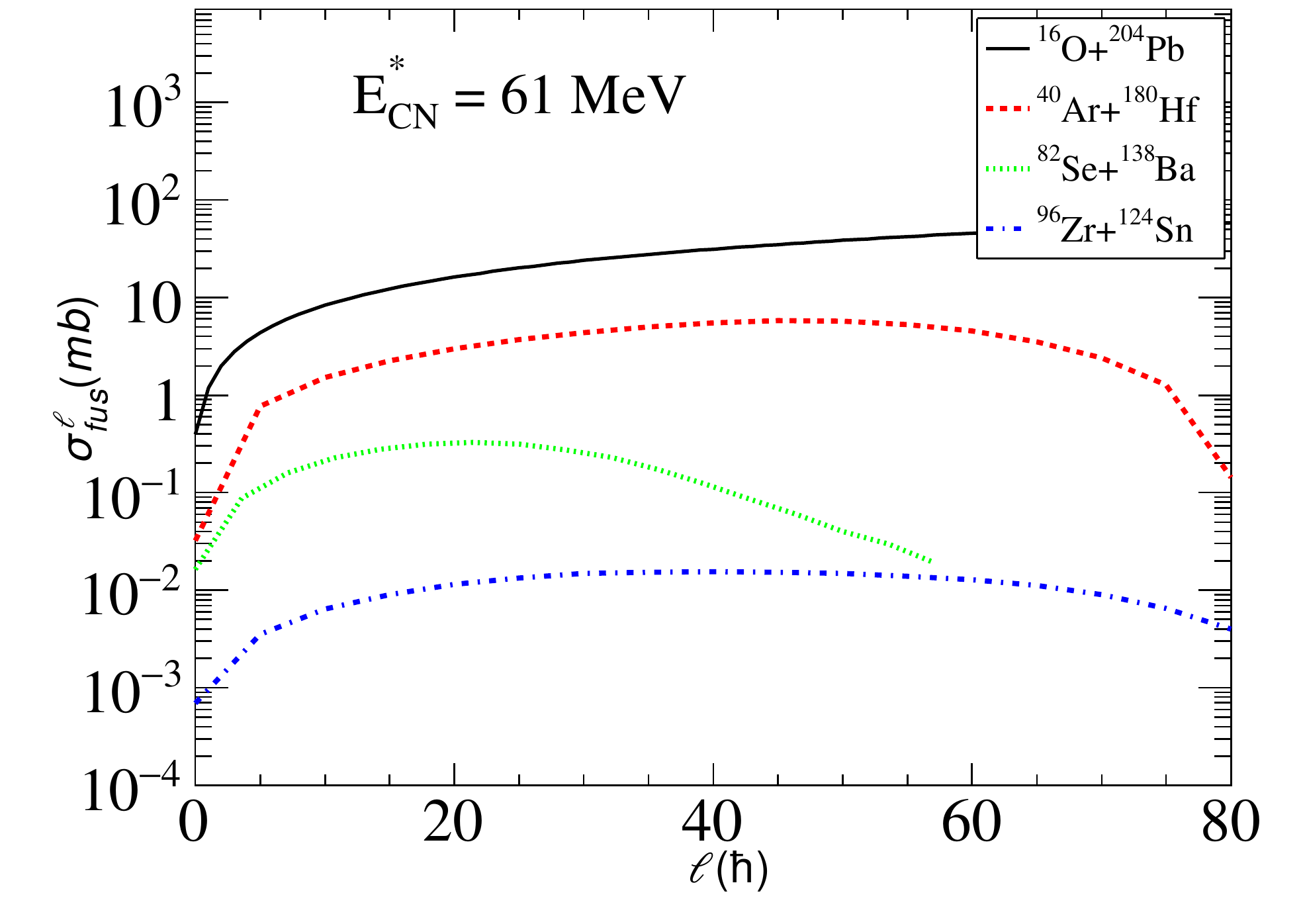}
\caption{(Color online) Partial fusion cross sections $\sigma_{\rm fus}^{\ell}$ as a function of angular momentum $\ell$ for the four considered reactions in the entrance channel leading to the same $^{220}$Th CN, at three fixed excitation energies of CN: $E^*_{\rm CN}$=35.5, 46, and 61 MeV (see insert for details).   \label{spin}}
\end{figure}

\begin{figure}
\centering
\includegraphics[scale=0.35]{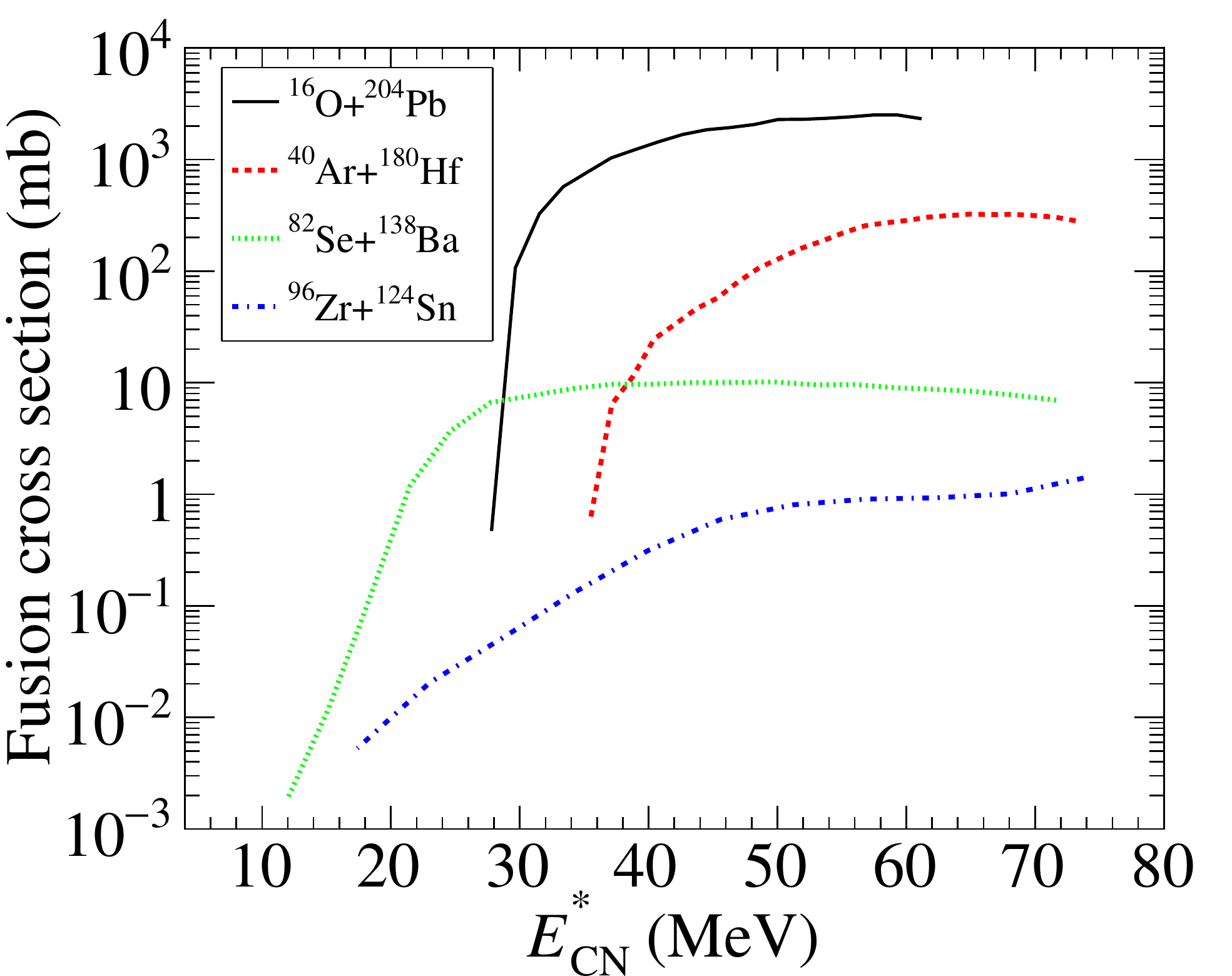}
\caption{(Color online) Fusion excitation functions for the  $^{16}$O+$^{204}$Pb (solid line),  $^{40}$Ar+$^{180}$Hf (dashed line), $^{82}$Se+$^{138}$Ba (dotted line)  and $^{96}$Zr+$^{124}$Sn (dash-dotted line) reactions. \label{fusion}}
\end{figure}

The role of the angular momentum distribution of the entrance channel on the capture and fusion cross sections and consequently on the evaporation residue products was in general discussed in  papers \cite{PRC91,NPA17}.
 The angular momentum distribution of the partial fusion cross section $\sigma_{\rm fus}^{\ell}$  versus $\ell$ for the four very different mass (charge) asymmetric reactions leading to the same $^{220}$Th CN formed at the same excitation energy $E^*_{\rm CN}$: 35.5, 46, and 61 MeV, respectively,
  is presented in Fig. \ref{spin}.
The choice of 35.5 MeV as the lower   $E^*_{\rm CN}$ value for which we can compare the effects of the angular momentum distribution for the complete set of the four considered reaction is due to the fact that for the $^{40}$Ar+$^{180}$Hf  reaction the threshold energy for the DNS formation (and consequently to reach the $^{220}$Th CN) is 35.5 MeV while for the other three  $^{16}$O+$^{204}$Pb, $^{82}$Se+$^{138}$Ba  and $^{96}$Zr+$^{124}$Sn reactions the threshold energy values are lower (see Fig. \ref{fusion}). Moreover, for the
  $^{40}$Ar+$^{180}$Hf  asymmetry reaction ($\eta$=0.64) the moment of inertia for dinuclear system $\mathcal{J}_{\rm DNS}$ is small, therefore, at low energy of about  $E^*_{\rm CN}=$35.5 MeV only collisions between  $^{40}$Ar and $^{180}$Hf with angular momentum $\ell\le20\hbar$ lead to capture, since at the corresponding energy of the center-of-mass $E_{\rm c.m.}$ entrance channel energy, the nucleus-nucleus interaction potential \cite{PRC91} $V(R,\ell)$ still forms a potential well up to $\ell=20\hbar$. Consequently, the partial fusion cross section $\sigma^\ell_{\rm fus}$ of the $^{220}$Th CN can be contributed from $\ell=0$ to 20~$\hbar$, while for $\ell>20\hbar$ the deep-inelastic collisions take place. Moreover, the comparison of the fusion excitation functions, calculated for the considered reactions (see Fig. \ref{fusion}) shows only for the   $^{40}$Ar+$^{180}$Hf  reaction a strong variation (more than 2 orders of magnitude) in the range of the excitation energy $E^*_{\rm CN}=$35.5 - 55 MeV while the trends for the  $^{16}$O+$^{204}$Pb  and $^{96}$Zr+$^{124}$Sn reactions are almost saturated or even slightly decreasing for the $^{82}$Se+$^{138}$Ba reaction.

 It is seen in the panels of Fig. \ref{spin} the relevant differences between the shapes of curves
    representing  the partial fusion excitation functions $\sigma_{\rm fus}^{\ell}$ for the four
    reactions
     leading to the same $^{220}$Th CN with the same excitation energy $E^*_{\rm CN}$ value and mainly between the different angular momentum $\ell$ intervals that cover these  $\sigma_{\rm fus}^{\ell}$ functions. Obviously, the different   yields
of the $\sigma_{\rm fus}^{\ell}$ functions are related to the different formations of the capture cross section for the four reactions and, consequently, to the  $\sigma_{\rm fus}^{\ell}$ fusion cross section values corresponding to the excitation energies of the compound nucleus  $E^*_{\rm CN}$ of 35.5, 46 and 61 MeV, respectively.  The specific shape of $\sigma_{\rm fus}^{\ell}$ and range of $\ell$ of each reaction determines the relevant differences in the type of the deexcitation  of  CN in the different reactions  leading to the same CN with the same excitation energy $E^*_{\rm CN}$.

 Our calculations
of each step of reaction to study the main properties of various processes have shown that:
\begin{enumerate}
\item The competition between the quasifission and complete fusion processes at the deexcitation of DNS is strongly sensitive to the characteristics of reactants in the entrance channel and then to the orbital angular momentum distribution of the excitation functions $\sigma^{\ell}_{\rm fus}$ and $\sigma^{\ell}_{\rm qf}$ of fusion and quasifission, respectively. The main role is played by the values of the $B_{\rm fus}^*$ intrinsic fusion barrier and $B_{\rm qf}$ quasifission barrier that are both sensitive to the angular momentum $\ell$ values.
\item The contributions of  the fast fission (ff) process and    compound nucleus formation are sensitive to the entrance channel through the angular momentum distribution characterizing the excitation functions of the related $\sigma^{\ell}_{\rm ff}$ and $\sigma^{\ell}_{\rm fus}$ cross sections.

\item The angular momentum distribution of the compound nucleus determines the competition between the deexcitation cascade of CN through the 
  evaporation of light particles (that can lead to the formation of ER nuclei with $Z_i$ and $A_i$ values not very different from the $Z$ 
  and $A$ numbers of CN)
  and its fission to two fragments with the charge $Z_i$ and and mass $A_i$ numbers near $Z/2$ and $A/2$, where $Z$ and $A$
   are the atomic and mass numbers of CN, respectively.

\end{enumerate}

 The probabilities of emission of light particles (n, p, and $\alpha$-particle) and fission processes
 are estimated by calculations of $\Gamma_n$, $\Gamma_p$,  $\Gamma_\alpha$, and $\Gamma_{\rm fis}$ widths, respectively,
  for each intermediate  excited nucleus formed during the deexcitation cascade starting from the CN.  The widths are determined by the corresponding 
  level densities which are function of the separation energies of emitted particles and the fission barrier.
 Moreover, the shell corrections in the fission barriers and the level densities $\rho$ are damped with functions depending on the excitation energies $E^*$ of nuclei and orbital angular momentum $\ell$.
The fission and  particle decay widths are calculated by considering the collective enhancement coefficients
 in the collective level density in order to correctly take into account in addition to the intrinsic excitations also the rotational and vibrational states by the collective enhancement factors~\cite{NPA17,darrigo94}.
Therefore, while the adiabatic approach for the estimation of the collective enhancement in the level density is acceptable only at very low excitation energies of few MeV, at higher  $E^*_{\rm CN}$ excitation energies it is necessary to use the non-adiabatic approach for the correct estimation of the collective level density $\rho_{\rm coll}^{\rm non-adiab.}(E^*, J)$. We use a damping function $q(E^*, \beta)$ with the aim to account the coupling of the collective to the intrinsic degrees of freedom due to the nuclear viscosity because it is rather unlikely that at high energies the adiabatic assumption still holds.
For details see Appendix B of \cite{NPA17} regarding the intrinsic and collective level density determinations, where we also show the sensitivity of the model on final reaction products by using mass asymmetric and almost symmetric reactants in the entrance channels.

The method of our study for the deexcitation of CN does not use free parameters that can be changed for each reaction, but it is rigorously applied to all reactions and for any explored excitation energy range.

\section{Results on the reactions leading to the $^{220}$Th CN}

We study some properties of the deexcitation of the  $^{220}$Th CN formed by four very different mass  (charge)  asymmetric and almost symmetric reactions in the entrance channel ( $^{16}$O+$^{204}$Pb,  $^{40}$Ar+$^{180}$Hf, $^{82}$Se+$^{138}$Ba and $^{96}$Zr+$^{124}$Sn) in a wide range of excitation energy
$E^*_{\rm CN}$.

\begin{figure}
\centering
\includegraphics[scale=0.35]{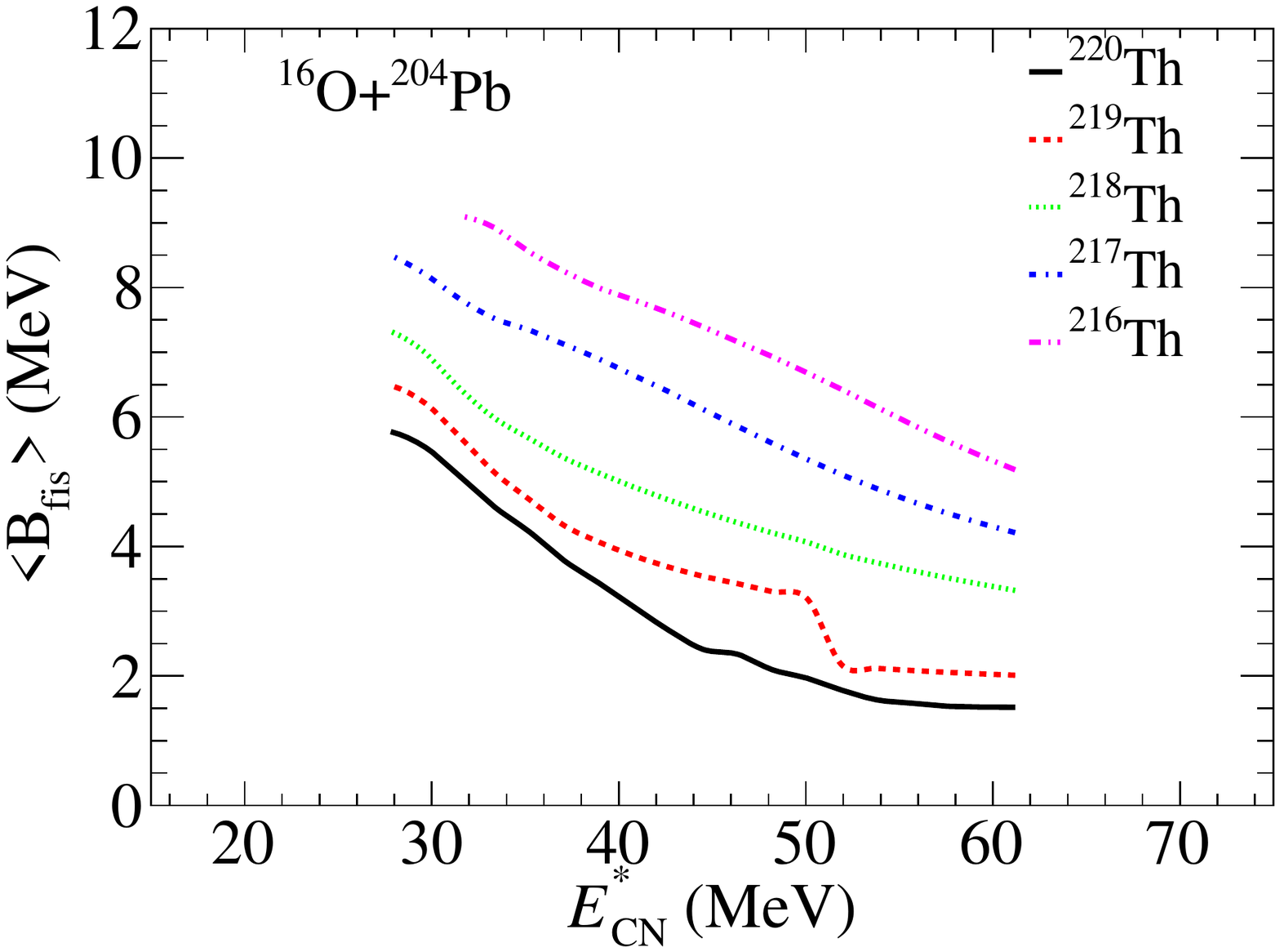}
\put(-150, 50){a)}\\
\includegraphics[scale=0.35]{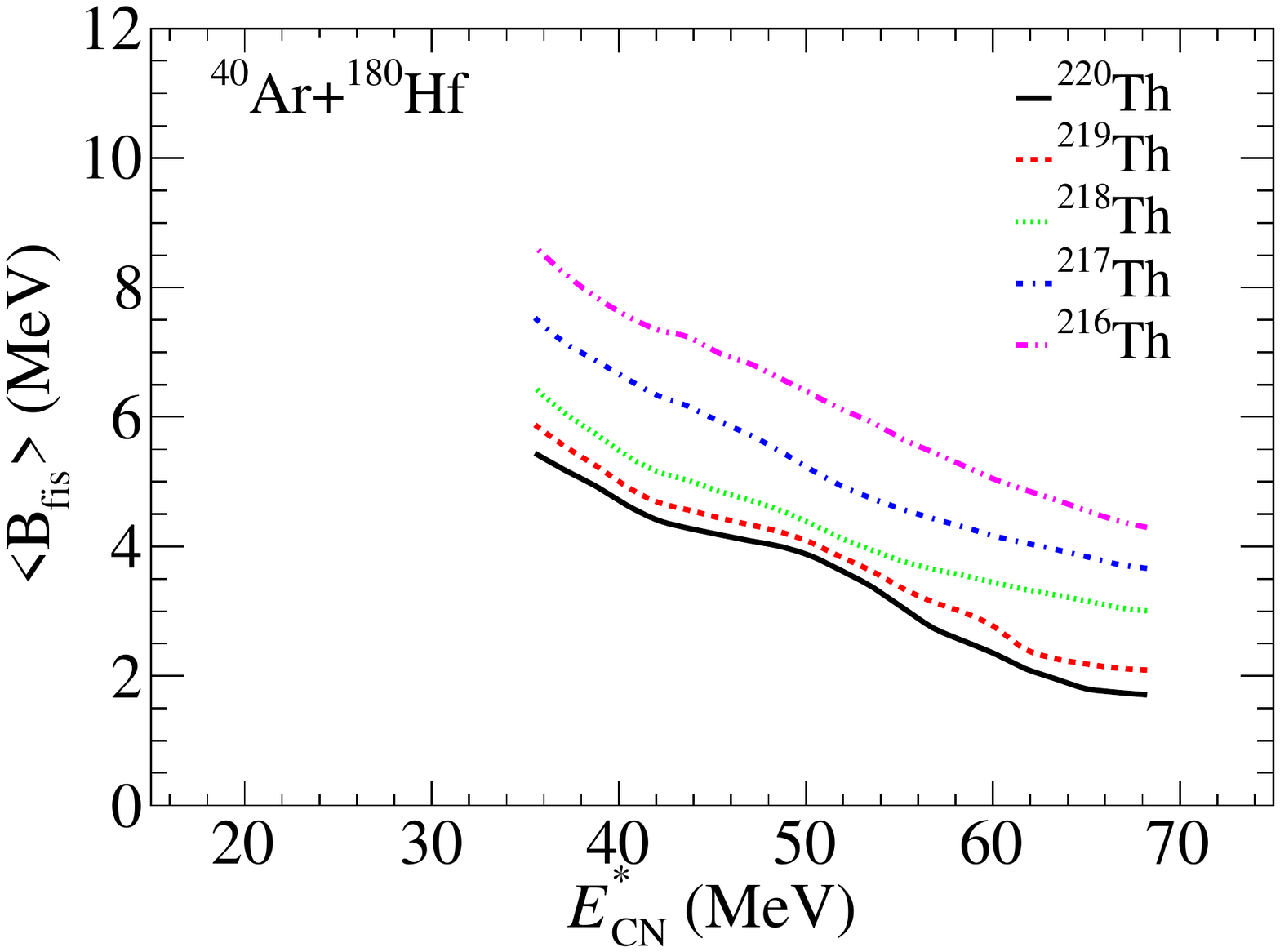}\put(-150, 50){b)}\\
\includegraphics[scale=0.35]{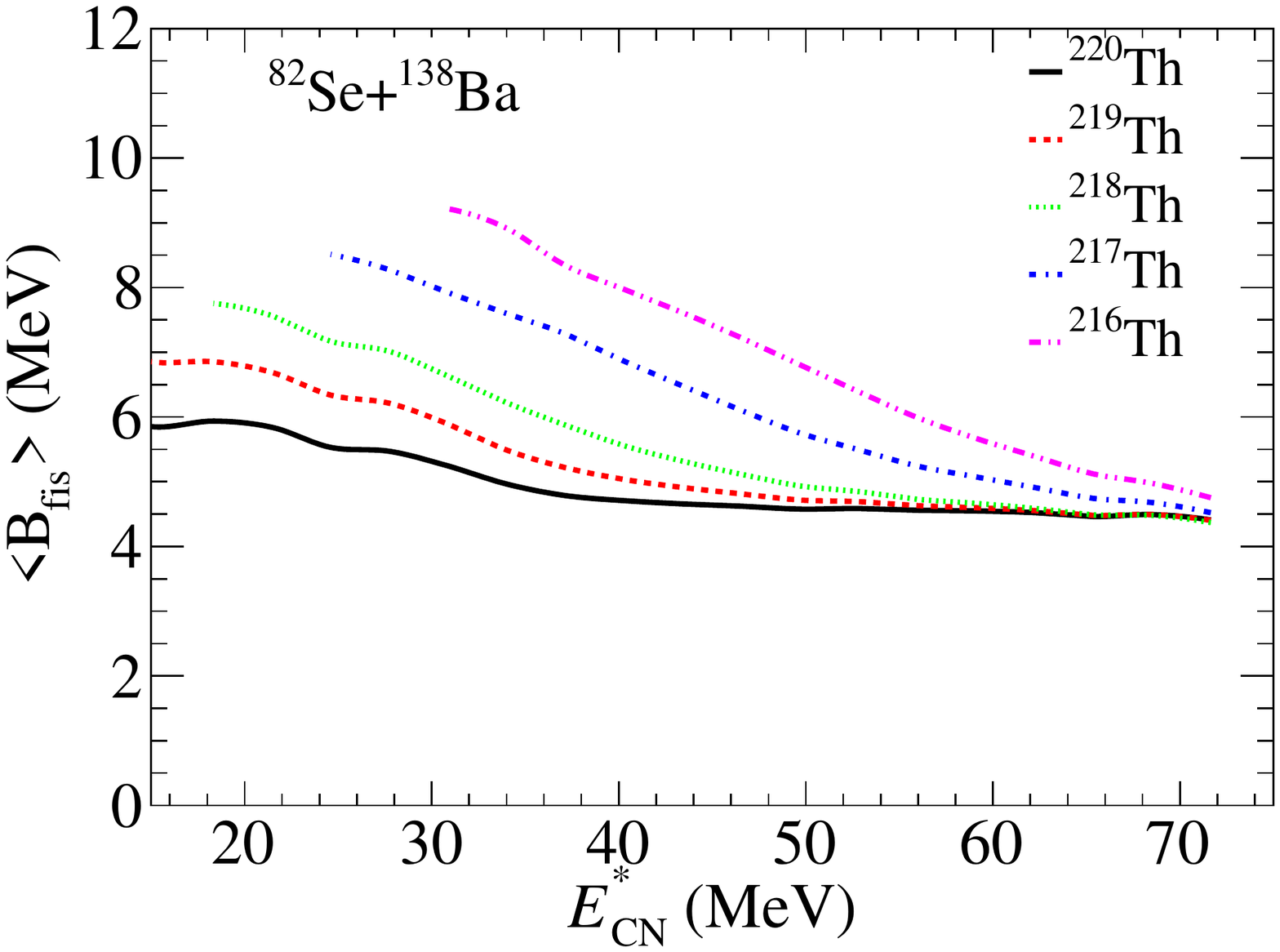}
\put(-150, 50){c)}\\
\includegraphics[scale=0.35]{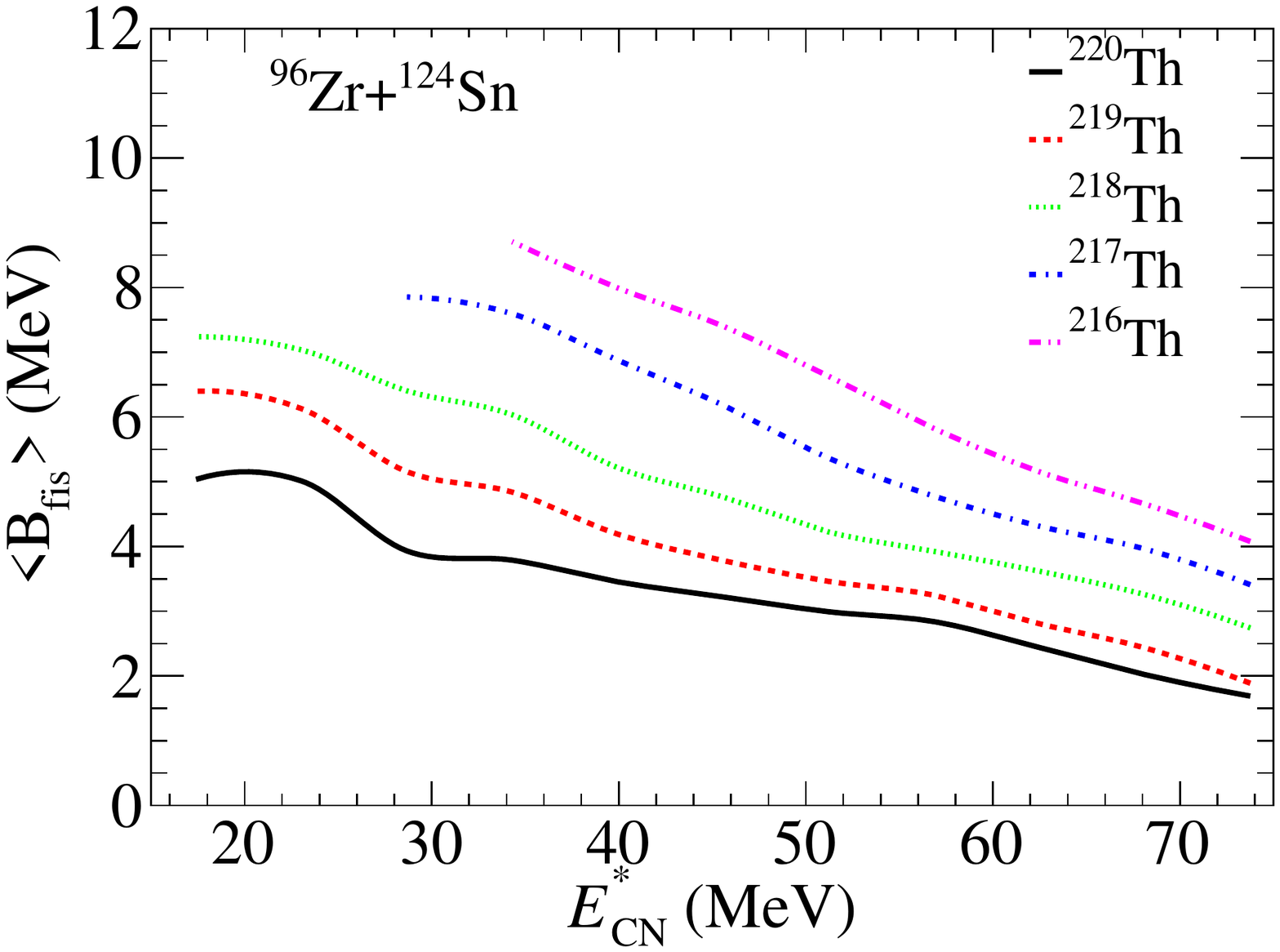}
\put(-150, 50){d)}
\caption{(Color online) The effective fission barrier $<B_{\rm fis}>$ for the excited $^{220}$Th$^*$ CN and some of its isotopes $^{219,218,217,216}$Th$^*$ formed after the successive neutron evaporation processes by the $^{16}$O+$^{204}$Pb,  $^{40}$Ar+$^{180}$Hf, $^{82}$Se+$^{138}$Ba and $^{96}$Zr+$^{124}$Sn reactions presented in panels a), b), c) and d), respectively.  \label{figbfis}}
\end{figure}

In  panel a) of Fig. \ref{figbfis}, we present the effective fission barrier $<B_{\rm fis}>$ versus the excitation energy $E^*_{\rm CN}$ for
 the $^{220}$Th$^*$ (full line) obtained in the $^{16}$O+$^{204}$Pb very  mass asymmetric reaction.  The same quantity calculated
 for the $^{219}$Th$^*$ (dashed line), $^{218}$Th$^*$ (dotted line),  $^{217}$Th$^*$ (dash-dotted line), and $^{216}$Th$^*$ (dash-double dotted line) after emission of one, two, three and four neutrons  from $^{220}$Th$^*$, respectively, are presented there.

The effective fission barrier $<B_{\rm fis}>$ value of CN with excited energy $E^*_{\rm CN}$ and any intermediate excited nucleus with energy $E^*$ is obtained as
the weighted average of  $B_{\rm fis}(\ell,T)$  by the partial $\sigma^{\ell}_{\rm fus}$:

\begin{equation}
B_{\rm fis}(T) = \frac{\sum_{\ell=0}^{\ell_{\rm d}} \sigma^{\ell}_{\rm fus}B_{\rm fis}(\ell,T)}{\sum_{\ell=0}^{\ell_{\rm d}}\sigma^{\ell}_{\rm fus}}
\label{averageBf}
\end{equation}

where the fission barrier $B_{\rm fis}(\ell,T)$ is:
\begin{equation}
B_{\rm fis}(\ell,T) =  B_{\rm fis}^{\rm m} - h(T)q(\ell)\delta W.
\label{bfis}
\end{equation}
In formula (\ref{bfis}), $T=\sqrt{E^*/a}$ represents the nuclear temperature where $a$ is the intrinsic level density parameter, and   $B_{\rm fis}^{\rm m}$ is   the part of the rotating liquid drop model contribution to the fission barrier;
 while $h(T)$ and $q(\ell)$ represent the damping functions of the nuclear shell correction $\delta W$ by the increase of the excitation energy $E^*$ and angular momentum $\ell$, respectively,
\begin{eqnarray}
h(T)& = &\{ 1 + \exp [(T-T_{0})/ d]\}^{-1}\label{damt}\\
q(\ell)& =& \{ 1 + \exp [(\ell-\ell_{1/2})/\Delta \ell]\}^{-1}.
\label{daml}
\end{eqnarray}
In equation (\ref{damt}), $d= 0.3$~MeV is the rate of  washing out the shell corrections with
the temperature, and $T_0=1.16$~MeV is the value at which the damping
factor $h(T)$
is reduced by 1/2. Analogously, in Eq. (\ref{daml}), $\Delta \ell
=3\hbar$ is the rate of washing out the shell corrections with the angular momentum, and $\ell_{1/2}
=20\hbar$ is the value
at which the damping factor $q(\ell)$ is reduced by 1/2.
It is useful to note that the values of parameters $d$, $T_{0}$, $\ell_{1/2}$ and $\Delta \ell$ used in the damping functions $h(T)$ and q($\ell$)
are not changed in the study of heavy ion reactions leading to heavy and superheavy nuclei. The parameters $d=0.3$~MeV and $T_0$=1.16 MeV in the damping function of the nuclear temperature were obtained by investigation of a very wide set of heavy ion reactions. The values of parameters $\ell_{1/2}=20\hbar$ and $\Delta\ell=3\hbar$ reduce the $q(\ell)$ function from 0.9 to 0.1 in the 12-26$\hbar$ interval confirming the important role of the  $q(\ell)$ damping function in determination of the effective fission barrier $<B_{\rm fis}(\ell,T)>$, as explained in Fig. 18 of Appendix B of paper \cite{NPA17}.

In our calculation, the intrinsic level density parameter $a$ is especially tailored to account for the shell effects in the level density \cite{NPA17}
\begin{equation}
a(E^*)= \tilde{a}\left\{ 1+\delta W \left[\frac{1-exp(-\gamma E^*)}{E^*}\right]     \right\}
\label{level_density_a}
\end{equation}
where  $\gamma=$0.0064 MeV$^{-1}$ is the parameter which accounts for the rate at which shell effects wash out with
excitation energy for neutron or other light particle emission.
The general expression (\ref{level_density_a}) works well also for deformed prolate or oblate nuclei.
Physically, the disappearance of the shell effects with $E^{*}$ excitation energy may be seen as a rearrangement of the shell-model orbitals
in such a way that the shell gap between orbitals close to the Fermi energy vanishes.
In order to determine the $a_{\rm fis}$ level density parameter in the
fission channel we use the relation $a_{fis}(E^*) =a_n(E^*)\times r(E^*)$ found in \cite{darrigo94} where $r(E^{*})$ is given by the relation
\begin{equation}
r(E^*)=\frac{\left[exp(-\gamma_{fis} E^*) - \left(1+\frac{E^*}{\delta W}\right) \right]}{\left[exp(-\gamma E^*) - \left(1+\frac{E^*}{\delta W}\right) \right]}
\label{ratio_a_parameters}
\end{equation}
with $\gamma_{fis}=0.024$~MeV$^{-1}$.

In our code, the fission and particle decay widths $\Gamma_{\rm fis}$ and $\Gamma_{\rm n,p,\alpha}$ are calculated by the formulas B1 and B2, respectively, reported in Appendix B of paper~\cite{NPA17}.

Similarly to what was done in panel a), we present in panels b), c), and d) of Fig. \ref{figbfis} the excitation functions of $<B_{\rm fis}>$ for the $^{40}$Ar+$^{180}$Hf asymmetric reaction, $^{82}$Se+$^{138}$Ba almost symmetric reaction, and  $^{96}$Zr+$^{124}$Sn  symmetric reaction, respectively.
\begin{figure}[hbt]
\centering
\includegraphics[scale=0.32]{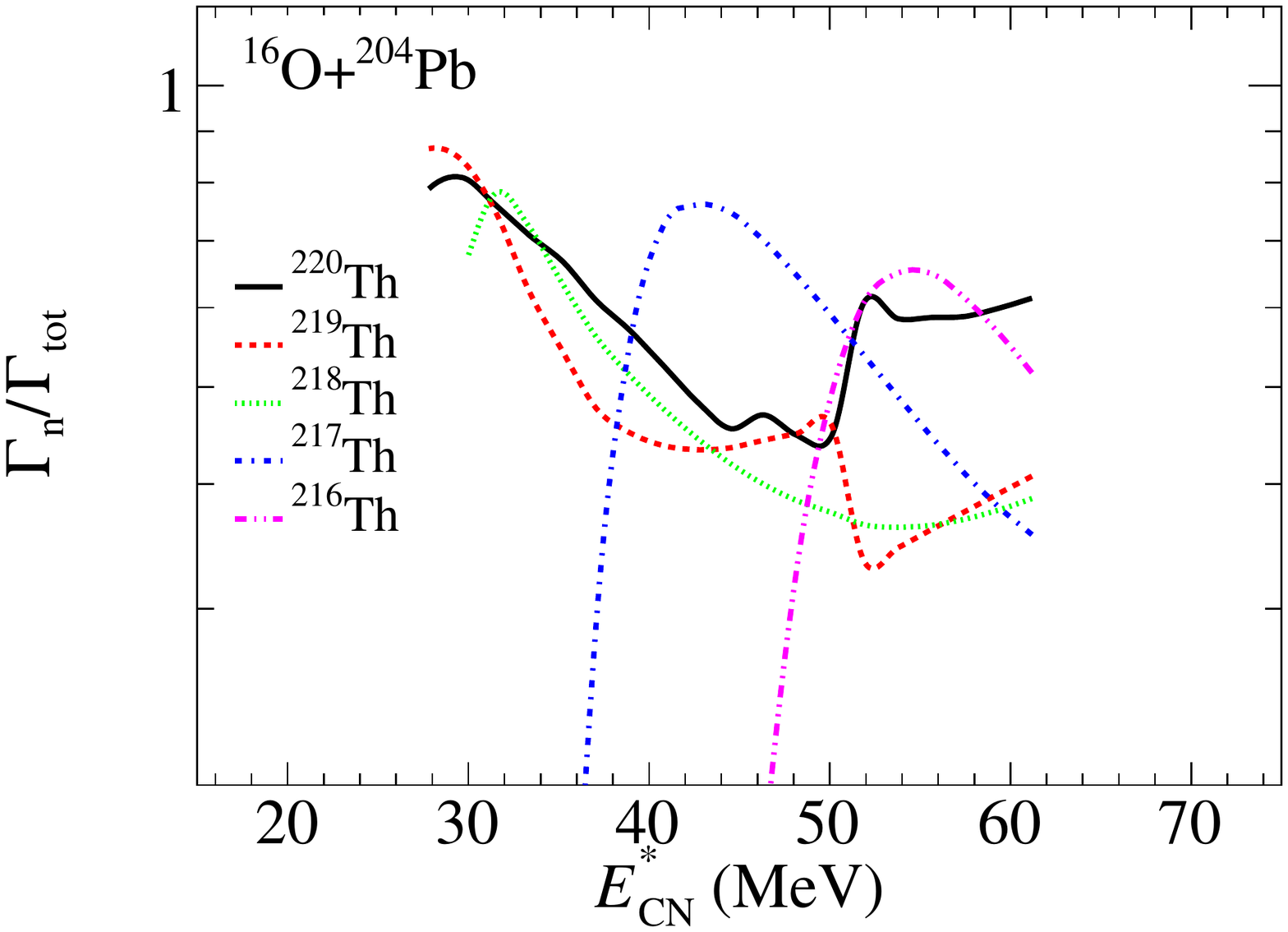}
\put(-140, 40){a)}\\
\includegraphics[scale=0.32]{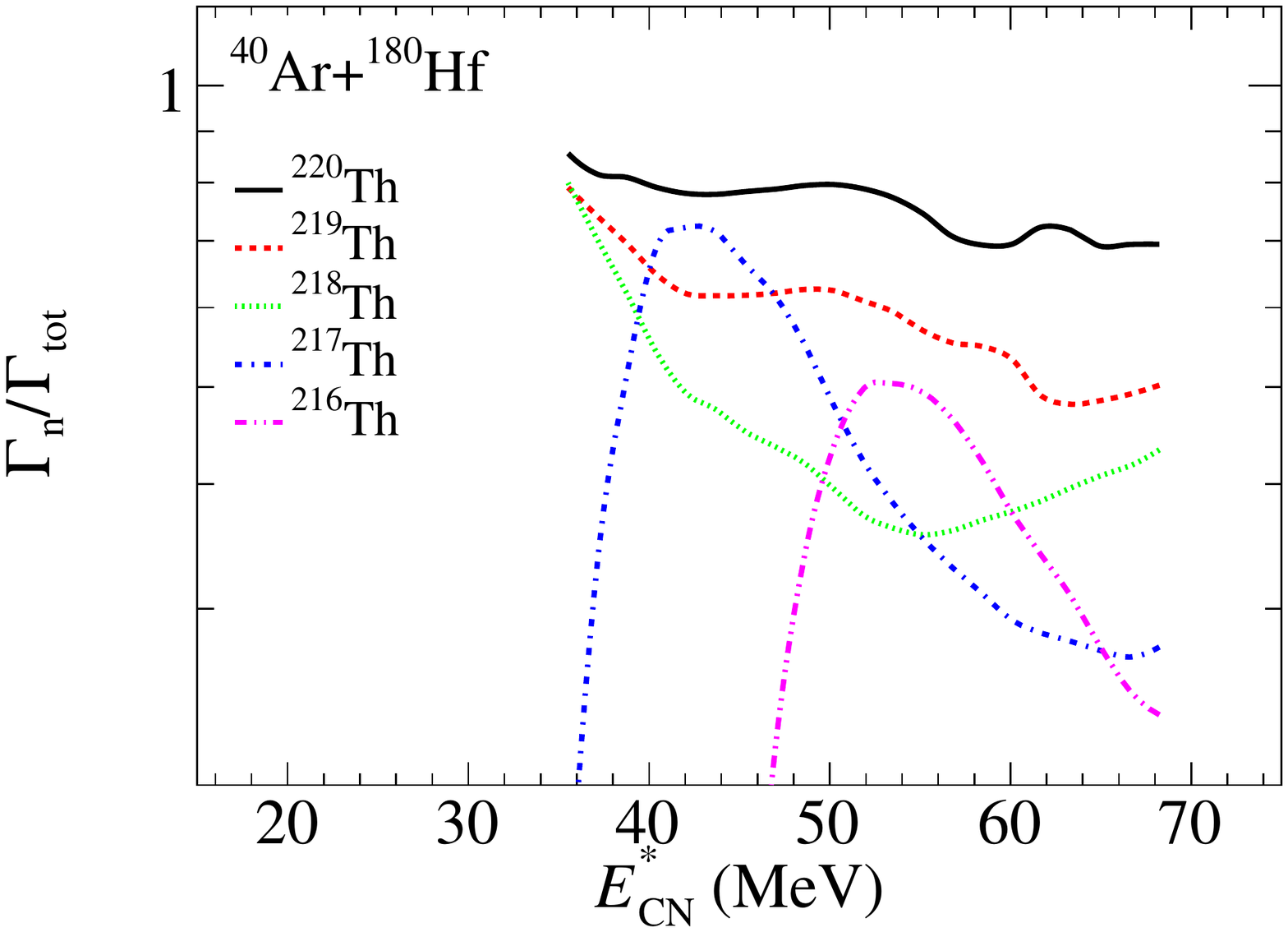}\put(-140, 40){b)}\\
\includegraphics[scale=0.32]{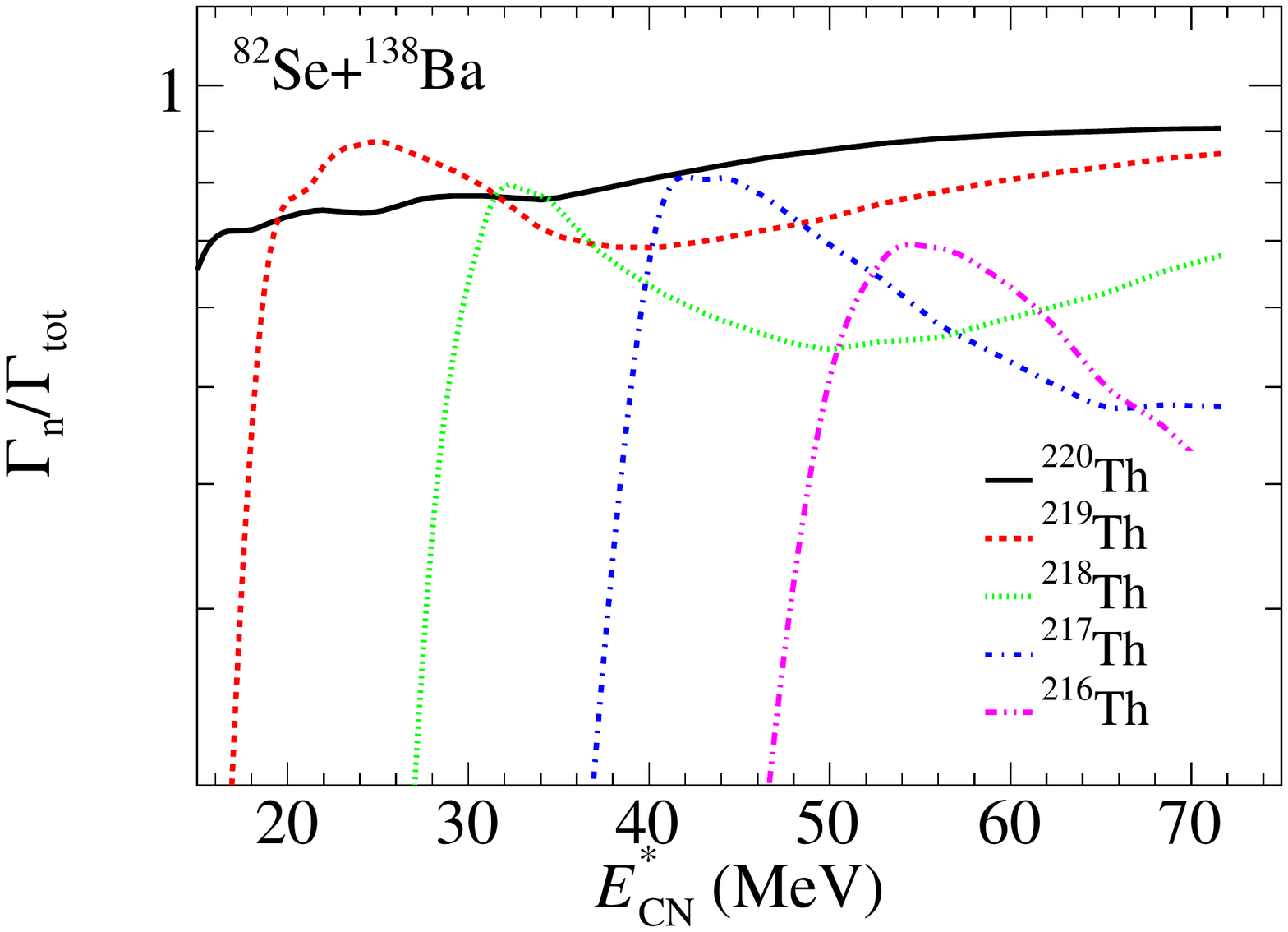}
\put(-143, 50){c)}\\
\includegraphics[scale=0.32]{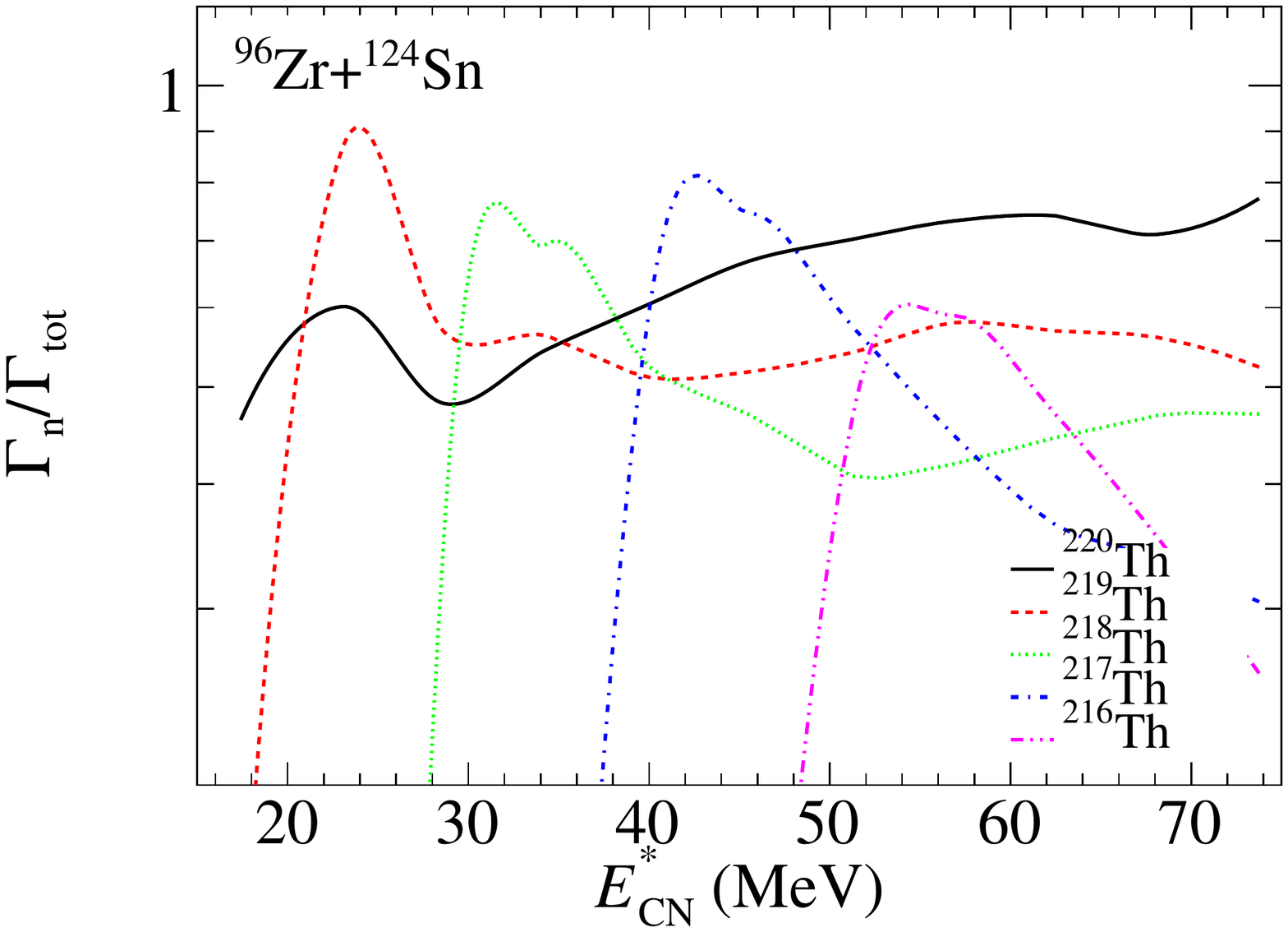}
\put(-145, 40){d)}\vspace{0.2cm}
\caption{(Color online) The excitation functions of the $\Gamma_{\rm n}/\Gamma_{\rm tot}$ neutron emission probability at deexcitation of the $^{220}$Th$^*$ CN and   $^{219,218,217,216}$Th$^*$ intermediate excited nuclei for the reactions and panels as indicated in Fig. \ref{figbfis}.  \label{figgngt}}
\end{figure}

Moreover,  the panels a), b), c), and d) of Fig. \ref{figgngt}  present the  branching ratio of the
calculated  neutron emission width results of excitation functions
$\Gamma_{\rm n}/\Gamma_{\rm tot}$  from the excited 
$^{220}$Th$^*$, $^{219}$Th$^*$, $^{218}$Th$^*$, $^{217}$Th$^*$, and $^{216}$Th$^*$  nuclei for the  very different entrance channels
 $^{16}$O+$^{204}$Pb,  $^{40}$Ar+$^{180}$Hf, $^{82}$Se+$^{138}$Ba, and $^{96}$Zr+$^{124}$Sn, respectively.
\begin{figure}[h]
\centering\vspace{-1cm}
\includegraphics[scale=0.32]{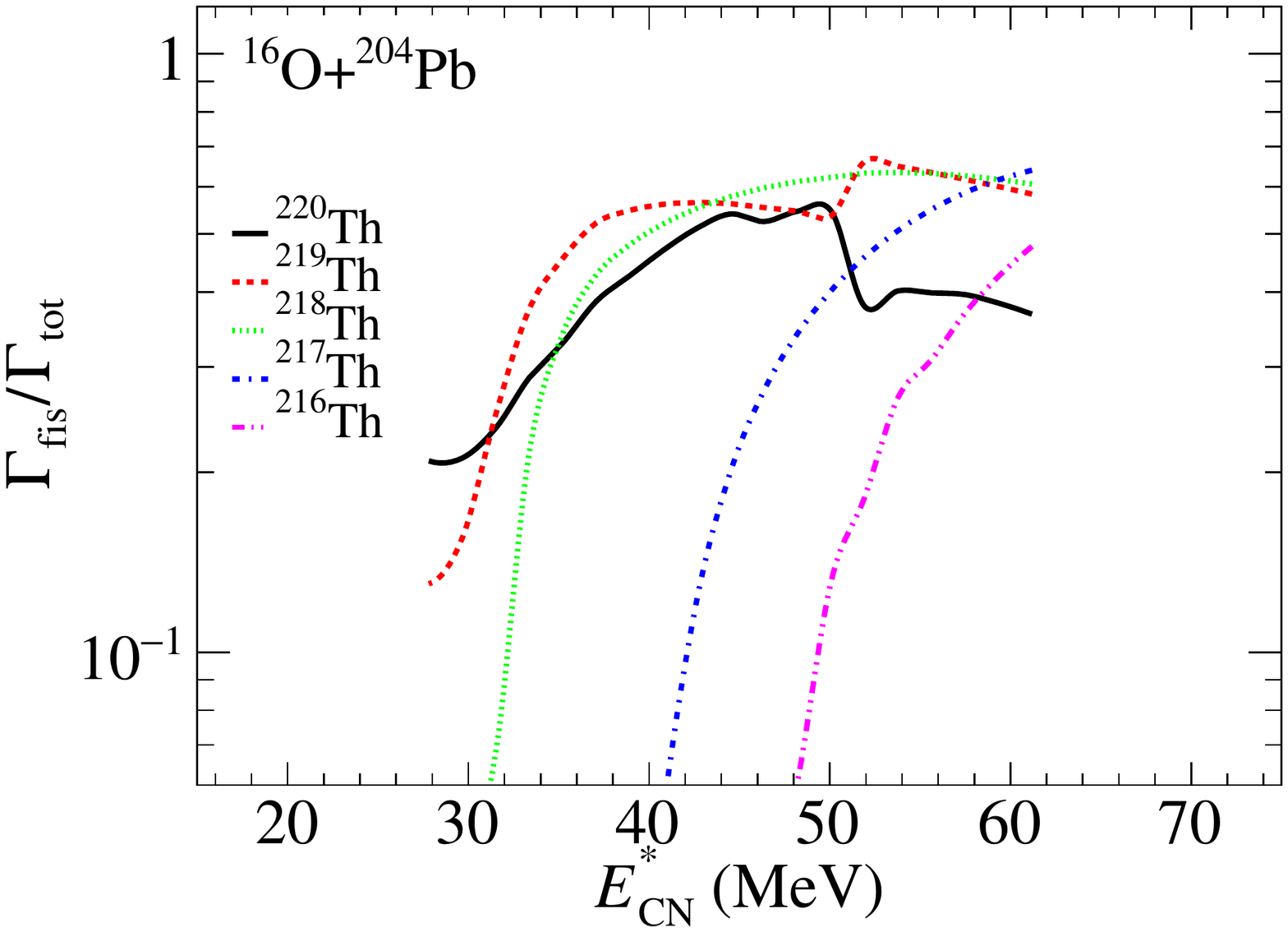}
\put(-145, 50){a)}\\
\includegraphics[scale=0.32]{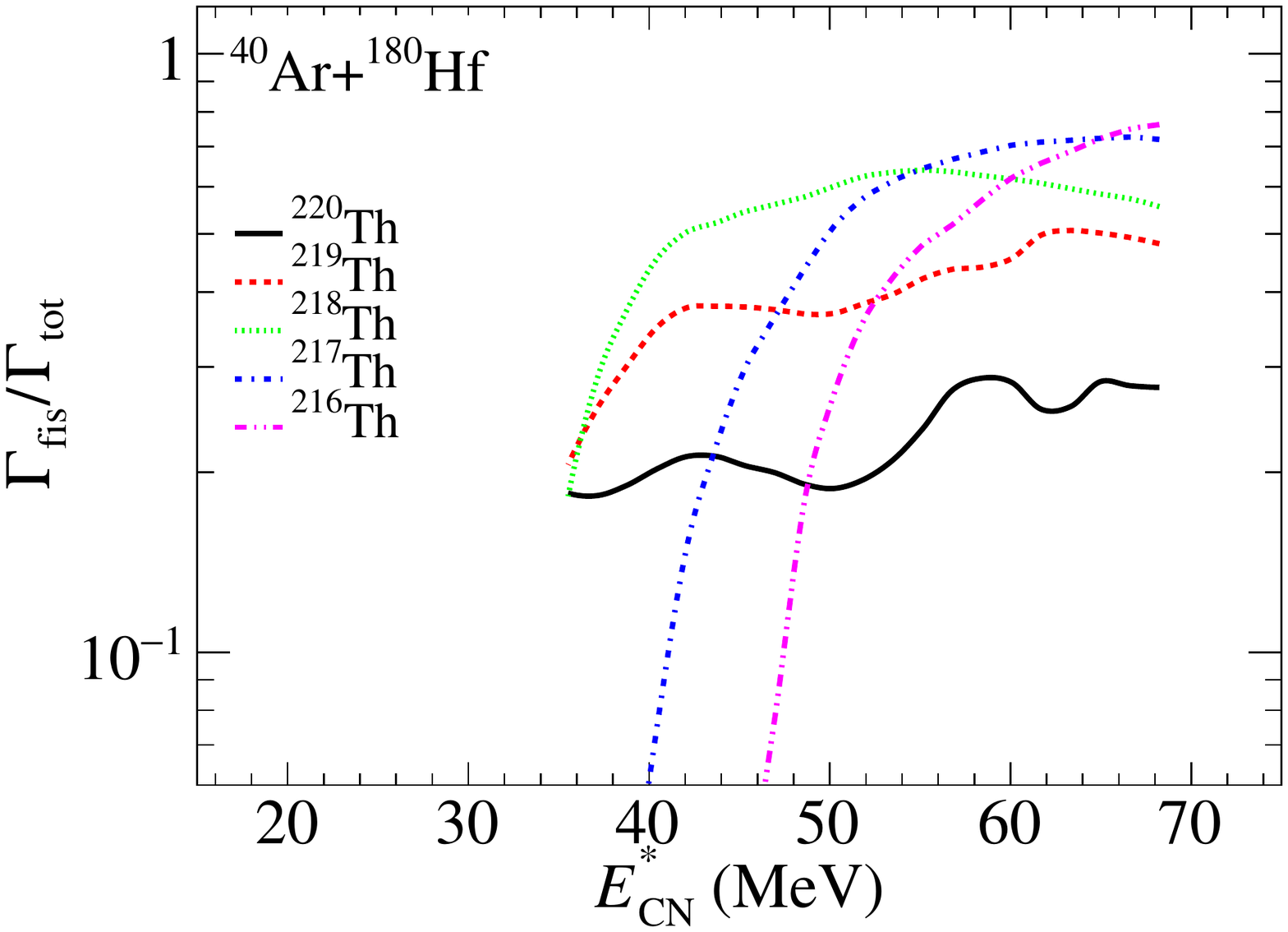}\put(-140,50){b)}\\
\includegraphics[scale=0.33]{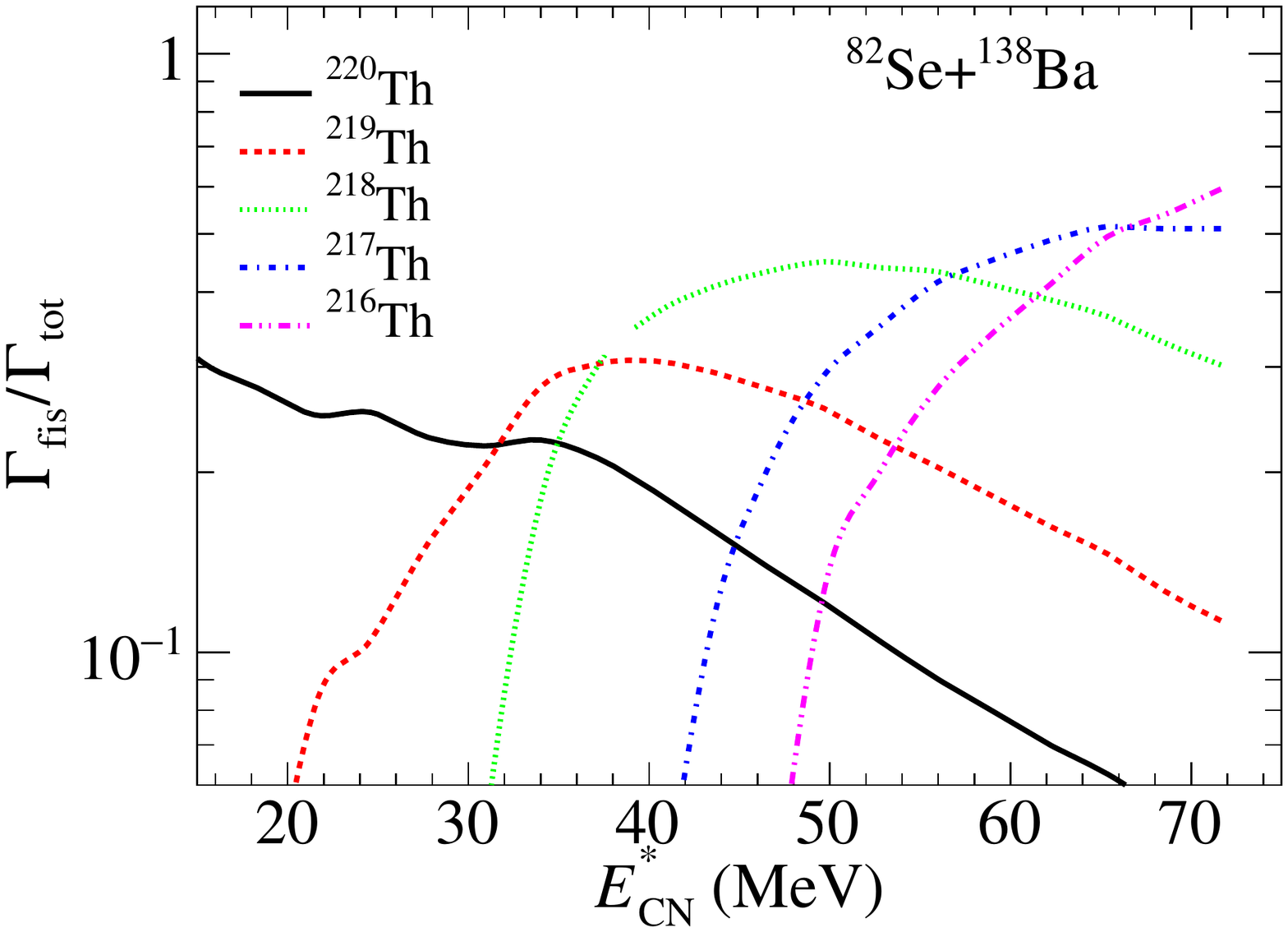}
\put(-145, 50){c)}\\
\includegraphics[scale=0.32]{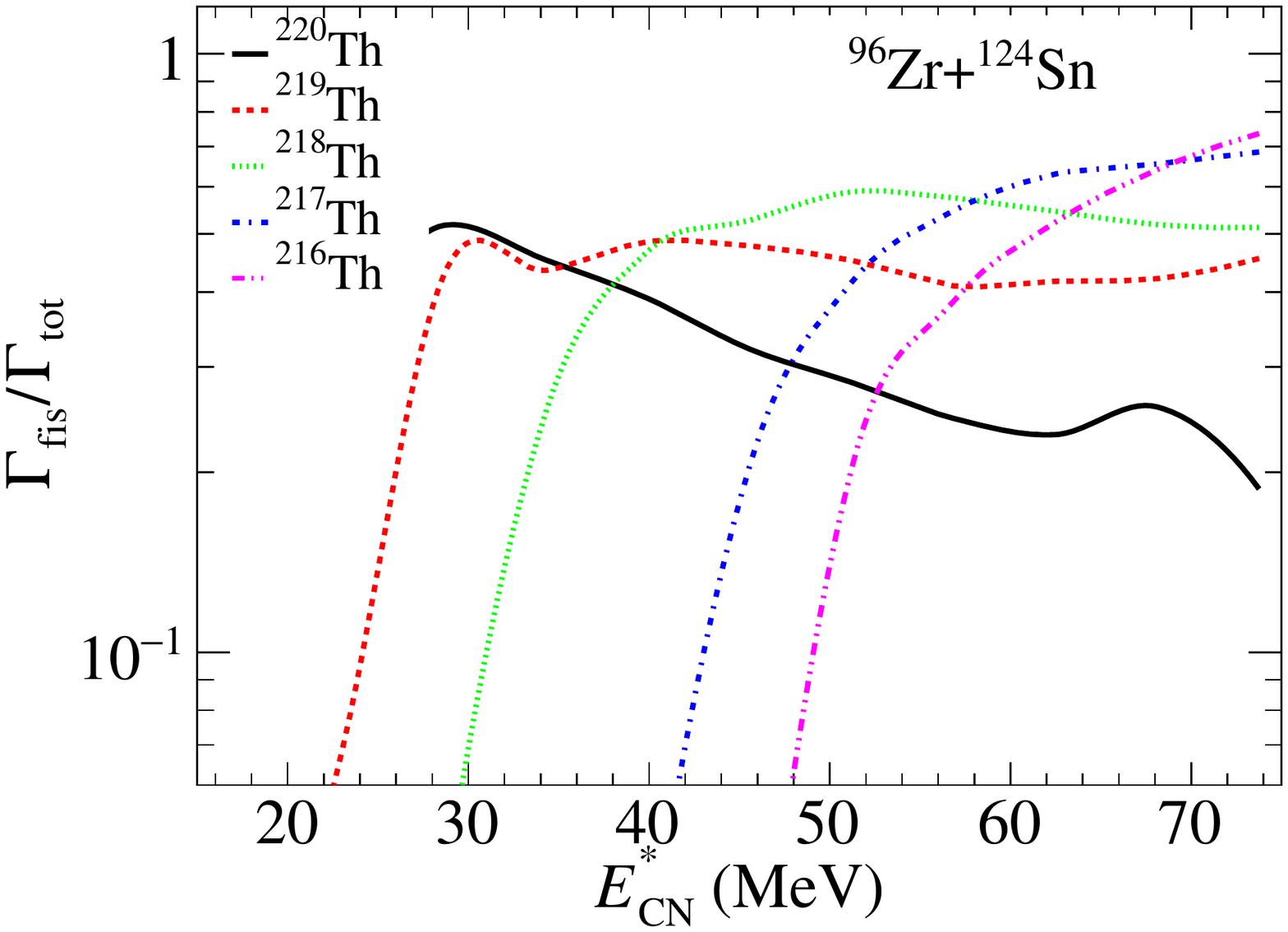}
\put(-140, 50){d)}
\caption{(Color online) As Fig. \ref{figgngt} but for the fission probability $\Gamma_{\rm fis}/\Gamma_{\rm tot}$. \label{figgfgt}}
\end{figure}
Analogously, in panels a), b), c), and d) of Fig. \ref{figgfgt} are reported the results of $\Gamma_{\rm fis}/\Gamma_{\rm tot}$ excitation functions representing the fission probabilities from the $^{220}$Th$^*$, $^{219}$Th$^*$, $^{218}$Th$^*$, $^{217}$Th$^*$, and $^{216}$Th$^*$ excited nuclei for the above-mentioned reactions.

 Figs. \ref{figbfis}, \ref{figgngt}, and \ref{figgfgt} show the trends of values of the $<B_{\rm fis}>$, $\Gamma_{\rm n}/\Gamma_{\rm tot}$, 
and $\Gamma_{\rm fis}/\Gamma_{\rm tot}$, respectively, as functions of $E^*_{\rm CN}$ for the 
 $^{220}$Th$^*$, $^{219}$Th$^*$, $^{218}$Th$^*$, $^{217}$Th$^*$, and $^{216}$Th$^*$ excited nuclei along the five steps of the deexcitation cascade of CN where the neutron evaporation from compound and other intermediate excited nuclei occurs.

In table \ref{tabfis} are reported the values of the effective fission barrier
$<B_{\rm fis}>$  calculated for the four investigated reactions at three fixed excitation energy of compound 35.5, 46 and 61 MeV, respectively, for the excited nuclei  $^{220,219,218,217,216}$Th$^*$ along the deexcitation cascade of CN after neutron emission only.
\begin{table}
\caption{Effective fission barrier $<B_{\rm fis}>$ values for the considered reactions at 35.5, 46, and 61 MeV of excitation energy of CN $E^*_{\rm CN}$ for the excited nuclei $^{220,219,218,217,216}$Th$^*$.\label{tabfis}}
\begin{tabular}{ |c||c|c|c|c|c|  }
 \hline
 \multicolumn{6}{|c|}{$<B_{\rm fis}>$ (MeV) at $E^*_{\rm CN}$=35.5 MeV} \\
 \hline
  & $^{220}$Th$^*$ &$^{219}$Th$^*$&$^{218}$Th$^*$& $^{217}$Th$^*$&$^{216}$Th$^*$\\
 \hline
$^{16}$O+$^{204}$Pb & 4.16 & 4.66 & 5.62 & 7.31 & 8.50\\
 \hline
$^{40}$Ar+$^{180}$Hf & 5.44 & 5.88 & 6.46 & 7.53 & 8.64\\
 \hline
 $^{82}$Se+$^{138}$Ba & 4.88 & 5.36 & 6.06 & 7.45 & 8.64\\
 \hline
 $^{96}$Zr+$^{124}$Sn & 3.72 & 4.70 & 5.86 & 7.44 & 8.56\\
 \hline
\hline

 \multicolumn{6}{|c|}{$<B_{\rm fis}>$ (MeV) at $E^*_{\rm CN}$=46 MeV} \\
 \hline
  & $^{220}$Th$^*$ &$^{219}$Th$^*$&$^{218}$Th$^*$& $^{217}$Th$^*$&$^{216}$Th$^*$\\
 \hline
$^{16}$O+$^{204}$Pb & 2.35 & 3.45 & 4.40 & 5.91 & 7.21\\
 \hline
 $^{40}$Ar+$^{180}$Hf &4.14 & 4.40 & 4.80 & 5.85 & 6.92\\
 \hline
 $^{82}$Se+$^{138}$Ba & 4.63 & 4.83 & 5.15 & 6.17 & 7.29\\
 \hline
 $^{96}$Zr+$^{124}$Sn & 3.20 & 3.76 & 4.72 & 6.12 & 7.35\\
 \hline
 \hline

 \multicolumn{6}{|c|}{$<B_{\rm fis}>$ (MeV) at $E^*_{\rm CN}$=61 MeV} \\
 \hline
  & $^{220}$Th$^*$ &$^{219}$Th$^*$&$^{218}$Th$^*$& $^{217}$Th$^*$&$^{216}$Th$^*$\\
 \hline
$^{16}$O+$^{204}$Pb & 1.52 & 2.01 & 3.33 & 4.23 & 5.21\\
 \hline
$^{40}$Ar+$^{180}$Hf & 2.21 & 2.56 & 3.38 & 4.10 & 4.95\\
 \hline
 $^{82}$Se+$^{138}$Ba & 4.53 & 4.57 & 4.62 & 4.97 & 5.50\\
 \hline
 $^{96}$Zr+$^{124}$Sn & 2.54 & 2.92 & 3.70 & 4.43 & 5.33\\
 \hline
\end{tabular}
\end{table}

These results are caused and determined by the different angular momentum distributions of the partial fusion cross section for the four different entrance channels, even when the excitation energy   $E^*_{\rm CN}$ is the same (see for example, Fig.~\ref{spin} at $E^*_{\rm CN}=35.5$, 46 and 61 MeV of the  excitation energies of the formed CN by the four considered reactions in the entrance channel).

Similar considerations can be made by observing
the values of neutron emission probabilities $\Gamma_{\rm n}/\Gamma_{\rm tot}$ reported in Fig. \ref{figgngt} and  the corresponding values of the fission probabilities $\Gamma_{\rm fis}/\Gamma_{\rm tot}$ reported in Fig. \ref{figgfgt}, at any fixed excitation energy value of $E^*_{\rm CN}$ when the $^{220}$Th CN is formed by the four very different reactions in the entrance channel. Since the effective fission barrier $<B_{\rm fis}>$  values are significantly different for the excited nuclei $^{220}$Th$^*$, $^{219}$Th$^*$, .... $^{216}$Th$^*$  along the deexcitation cascade at any fixed excitation energy value of the formed $^{220}$Th CN, also the competition between the value of the neutron emission probability $\Gamma_{\rm n}/\Gamma_{\rm tot}$  and the one of the corresponding   fission probability $\Gamma_{\rm fis}/\Gamma_{\rm tot}$  is different for any studied excited nucleus reached along the deexcitation cascade. This assertion and the related differences can be easily verified by comparing 
 the $\Gamma_{\rm n}/\Gamma_{\rm tot}$ values  in Fig. \ref{figgngt} (and 
  the $\Gamma_{\rm fis}/\Gamma_{\rm tot}$  values  in Fig. \ref{figgfgt}) obtained at any $E^*_{\rm CN}$ value 
  (see for example at 35.5, 46, and 61 MeV) when the same $^{220}$Th CN and other intermediate excited nuclei are formed by the four different considered reactions in the entrance channel.

As discussion about results concerning $<B_{\rm fis}>$, $\Gamma_{\rm n}/\Gamma_{\rm tot}$  and $\Gamma_{\rm fis}/\Gamma_{\rm tot}$ represented in Figs. \ref{figbfis}, \ref{figgngt}, and \ref{figgfgt}, respectively, we give the following details.
The trends of the lines  $<B_{\rm fis}>$ given in Fig. \ref{figbfis} for the four considered reactions represented in panels (a), (b), (c) and (d),
 respectively, are consistent with each other taking into account that for the two very asymmetric reactions (panels (a) and (b)) the fusion process is dominant in comparison with the quasifission one   due to low values of the intrinsic fusion barrier $B^*_{\rm fus}$~\cite{PRC91}, especially at high values of angular momentum $\ell$ during the evolution of DNS. Instead, for the two almost symmetric reactions (panels (c) and (d) of figure \ref{figbfis}) 
 the quasifission process is dominant in comparison with the fusion one due to low value of quasifission barrier $B_{\rm qf}$ \cite{PRC91}, especially at high values of $\ell$. But, at the deexcitation of CN the fission process is dominant for asymmetric reactions (panels (a) and (b)) in comparison with the evaporation process of light particles (n, p, and $\alpha$) especially at high $E^*_{\rm CN}$ values where $<B_{\rm fis}>$  strongly decreases. Differently, for the symmetric reactions (panels (c) and (d) of figure \ref{figbfis}) a large part of high momentum values  contributes to quasifission during the 
 DNS stage determining the CN formation characterized by a reduced range of angular momentum $\ell$.  In these cases represented in panels (c) and (d), the  $<B_{\rm fis}>$ values slowly decrease with the increase of the excitation energy $E^*_{\rm CN}$ since  the damping function on the shell correction by formula  (\ref{daml})-produces a smaller fade-out of the fission barrier $B_{\rm fis}$ in comparison with the cases of panels a) and b) representing the asymmetric reactions, for any excitation energy $E^*_{\rm CN}$. Consequently to the results presented in Fig. (\ref{figbfis}) and the above comments, for the asymmetric reactions the fission probability $\Gamma_{\rm fis}/\Gamma_{\rm tot}$ of $^{220}$Th CN versus $E^*_{\rm CN}$ in average increases (see solid lines in panels (a) and (b) of Fig. \ref{figgfgt}) and obviously the neutron emission probability $\Gamma_{\rm n}/\Gamma_{\rm tot}$ of excited $^{220}$Th CN in average decreases versus $E^*_{\rm CN}$ (see solid lines in panels (a) and (b) of Fig. \ref{figgngt}).
Instead, for the symmetric reactions represented in panels (c) and (d) of Fig. \ref{figgngt}, at a weak increase of the neutron emission probability $\Gamma_{\rm n}/\Gamma_{\rm tot}$ versus $E^*_{\rm CN}$ corresponds a decrease of the fission probability  $\Gamma_{\rm fis}/\Gamma_{\rm tot}$ versus $E^*_{\rm CN}$ (see solid lines in panels (c) and (d) of Fig. \ref{figgfgt}).

Moreover, in order to verify the effects that are induced on the types of evaporation residue nuclei formed along the deexcitation cascade of the same $^{220}$Th CN (reached with the same value of excitation energy $E^*_{\rm CN}$ by a very different mass asymmetric reactions in the entrance channel) we also calculated the ER yields obtained by the intermediate excited nuclei (with $Z<90$) after charged particle emissions (proton and $\alpha$) together with neutron emission, and the ER yields obtained by the excited nuclei after neutron emission only (with $Z=90$).  For this purpose, we present in table \ref{tabratio} the ratio between the ER production from excited nuclei with $Z<90$ and the ER production from excited nuclei with $Z=90$, when the   $^{220}$Th CN is formed at excitation energies of  35.5, 46, and 61 MeV by the very different reactions in the entrance channel:  $^{16}$O+$^{204}$Pb, $^{40}$Ar+$^{180}$Hf, $^{82}$Se+$^{138}$Ba, $^{96}$Zr+$^{124}$Sn.

\begin{table}[h]
\caption{Values of the ratio ER($Z<90$)/ER($Z=90$) for the four considered reactions at three fixed excitation energies of the compound nucleus 35.5, 46, and 61 MeV.\label{tabratio}}
\begin{tabular}{ |c||c|c|c|c|  }
 \hline
 \multicolumn{5}{|c|}{Values of the ratio ER($Z<90$)/ER($Z=90$)} \\
 \hline
$E^*_{\rm CN}$(MeV) & $^{16}$O+$^{204}$Pb & $^{40}$Ar+$^{180}$Hf & $^{82}$Se+$^{138}$Ba & $^{96}$Zr+$^{124}$Sn \\
 \hline
35.5  & 9.5 & 2.9 & 8.7 &3.3 \\
 \hline
 46  & 8.2 & 6.0 & 10.1 &8.7 \\
 \hline
 61  & 10.7 & 15.2 & 13.3 &19.4 \\
 \hline
\hline
\end{tabular}
\end{table}

By considering the obtained results of the ratio  ER($Z<90$)/ER($Z=90$) it is possible to conclude that:
\begin{itemize}
\item[i)] the contribution of the ER formation after charged and neutral particle emission with $Z<90$ is many times higher than ER formation after neutron emission only, at any excitation energy $E^*_{\rm CN}$ of the compound nucleus formation, and for any type of reaction in the entrance channel;
\item[ii)] the rate of the ER formation after charged particle emission generally increases with the increase of the excitation energy value of  $E^*_{\rm CN}$;
\item[iii)] at any excitation energy value of $E^*_{\rm CN}$, the ratio ER($Z<90$)/ER($Z=90$) is strongly sensitive to the entrance channel reaction starting from the first nuclear contact of reactants (forming the DNS) up to the complete evolution of the reaction mechanism that is specific for each reaction, and that ratio changes for the same considered reaction with the excitation energy value of  $E^*_{\rm CN}$.
\end{itemize}

Therefore, the results presented in table \ref{tabratio} and the consequent related analysis of reactions leading to the same  $^{220}$Th CN allow one to conclude that the rates of the reaction products (quasifission, fusion-fission, fast fission, and the types of evaporation residues formed after neutron and charged particles emissions) are strongly dependent on the reaction in the entrance channel and on the excitation energy $E^*_{\rm CN}$. Moreover, the deexcitation cascade of the same CN formed with the same excitation energy value $E^*_{\rm CN}$ is strongly sensitive to the reaction in the entrance channel to cause of the different angular momentum distribution $\sigma_{\rm fus}^{\ell}$ versus $\ell$ at formation of CN.

\begin{figure}
\centering
\includegraphics[scale=0.35]{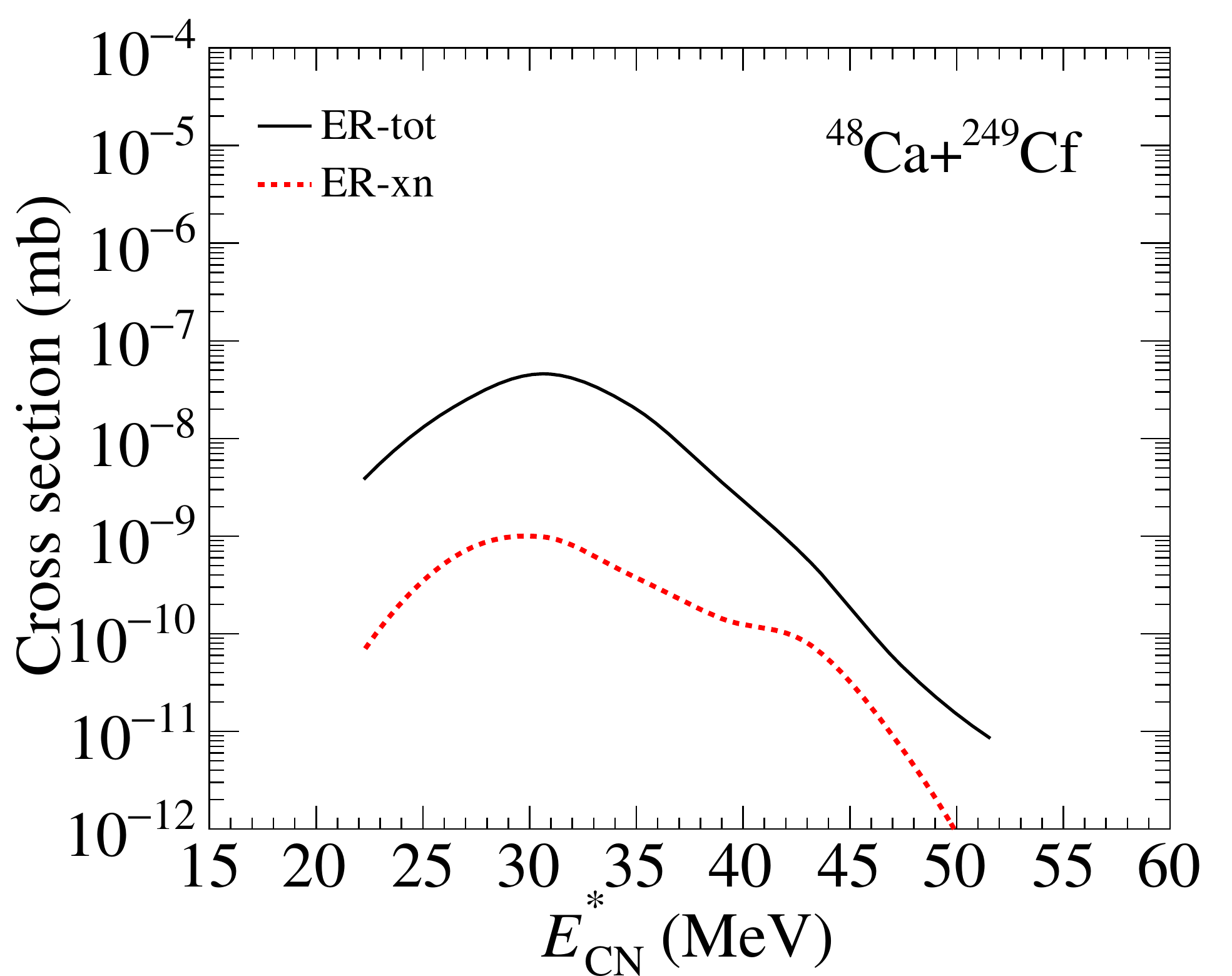}
\caption{(Color online) Evaporation residue cross section for the $^{48}$Ca+$^{249}$Cf reaction leading to the $^{297}$118 CN, for neutron emission only (dashed line) and by including in calculation also charged particles (solid line). \label{she}}
\end{figure}

The presented methodology and results on the excitation functions of evaporation residues formed after neutron and charged particles emissions can be useful to experimentalists in order to observe and measure with a better efficient rate the final products of heavy nuclei reactions leading to formation of superheavy compound nuclei. Since along the deexcitation cascade of superheavy CN it is possible for excited nuclei to evaporate charged particles ($\alpha$ and proton) together with neutrons,  and the formation of evaporation residues (ERs) after charged particle emission too is much larger in comparison with the ERs formation after neutron emission only (see Fig. \ref{she}), it is necessary in experiments to arrange facilities and methods useful to detect evaporation residues with $Z_{\rm ER}<Z_{\rm CN}$ within an interval of residue nuclei with atomic number $Z_{\rm i}$ ranging between  $Z_{\rm i}$=$Z_{\rm CN}-8$ and $Z_{\rm i}$=$Z_{\rm CN}$. The efficiency of detecting evaporation residues of superheavy nuclei and the possibility of determining the related cross sections are more favorable than those currently used for the identification of ERs after neutron emission only from the CN.

As an example we present in figure \ref{she} the comparison between the $\sigma_{\rm ER}$ cross sections versus $E^*_{\rm CN}$ obtained for the $^{48}$Ca+$^{249}$Cf reaction leading to the $^{297}$118 CN when the charged particles  are also considered in the evaporation (full line) and when the neutron emission only are considered (dashed line) along the deexcitation cascade of the compound nucleus. As one can see,  the cross section obtained by including the emission of charged particles overcomes the one obtained via neutron emission only by more than one order of magnitude.
The evaporation residue calculated for neutron emission and the comparison with experimental data was reported in our previous work \cite{aglio2012}. The theoretical group  \cite{adami16}  suggests the opportunity to use the charged emission channels for the synthesis of superheavy elements by making quantitative calculation, while others \cite{pirla1} affirm charged emission contribution to ERs being negligible in this kind of reaction. Therefore, accurate experimental investigation on the identification of the evaporation residue produced also by the emission of charged particles can give precious information to theoretical groups to improve the prediction power of their models.

\section{Conclusion}

The present investigation on heavy ion reactions with various mass  (charge)  asymmetry parameters points out the effects of the entrance channel on the CN formation and the consequent different ways of its deexcitation cascade even when the formed CN is characterized by the same Z and A values and has the same excitation energy $E^*_{\rm CN}$. The reason of this different way of deexcitation of CN  is due to the different orbital angular momentum distribution of reactants in the entrance channel to cause of: different mass  (charge)  asymmetry parameter
of reacting nuclei,  the specific shapes (oblate, prolate or spherical), and also  the eventual deformation parameters of these beam and target nuclei, even when the CN is formed with the same Z, A and $E^*_{\rm CN}$ values. Moreover, the effective fission barrier $<B_{\rm fis}>$  of each intermediate excited nucleus reached at each step of the deexcitation cascade of the same formed CN with the same $E^*_{\rm CN}$  but by various entrance channels, is affected by the various angular momentum distributions for the four considered reactions forming the $^{220}$Th CN; in addition,  the damping function of the fission barrier $B_{\rm fis}$  determines different effects on the deexcitation of CN to cause of the different ranges of  angular momentum $\ell$ covered by  the considered reactions. Therefore,
the competition between $\Gamma_{\rm n}/\Gamma_{\rm tot}$ and
$\Gamma_{\rm fis}/\Gamma_{\rm tot}$ for each considered
intermediate excited nucleus reached along the deexcitation cascade of CN is affected by the type of reaction in the entrance channel and the excitation energy $E^*_{\rm CN}$ too.
Furthermore, we can anticipate the information that at each $E^*_{\rm CN}$ value the formation of evaporation residue nuclei ERs --formed after neutrons emission only along the deexcitation cascade of CN and the ones formed by the charged emission too-- is strongly sensitive to the type of reaction in the entrance channel and to the complete reaction mechanism  also when it leads to the same CN for A, Z and $E^*_{\rm CN}$ values.

Therefore, by considering also the emission of charged particles proton and $\alpha$ together with neutron emission along the deexcitation cascade of the $^{220}$Th CN the results of calculation show that the total ER evaporation residue nuclei cross sections contributed by the intermediate excited nuclei after charged particle emission too (with $Z<$90) along the complete deexcitation cascade of CN are greater than the ER  yields obtained by the excited nuclei after neutron emission only (with $Z=$90).
This additional   result on the evaporation residue production further confirms that the deexcitation of the same CN with the same excitation energy  $E^*_{\rm CN}$ formed by various reactions with different mass (charge) asymmetry parameter is strongly sensitive to the effects of the entrance channel and also to the complete reaction mechanism that is specific for each reaction.
%

We have shown that the consistent description of the fade-out of the shell correction to the fission barrier with increasing temperature and angular momentum is very important for reliable estimation of the final reaction products. This result is of crucial importance for the synthesis of superheavy elements, since it extends the stabilizing effects of the shell structure to higher temperature, but this stabilizing effects, however, will be partial removed by the decrease of the shell correction with the increase of angular momentum. Moreover, the present result related to the determination of ERs nuclei formation after charged particle emission of CN indicating total evaporation residue cross sections much higher than the ER cross section after neutron emission only (see Fig. \ref{she}), suggest researchers in experiment of developing appropriate procedures useful for the identification of ER nuclei with $Z<Z_{\rm CN}$.

\section*{Acknowledgements}

A.K.N. thanks the Russian Foundation for Basic Research
for the partial support.


\end{document}